\documentclass[11pt]{article}

\usepackage{UF_FRED_paper_style}
\usepackage{CJKutf8}
\usepackage{booktabs}

\usepackage[blocks]{authblk}

\usepackage{lipsum}  

\title{Global Data Science Project for COVID-19}

\author[1]{Toyotaro Suzumura}
\author[2]{Dario Garcia-Gasulla}
\author[2]{Sergio Alvarez Napagao}
\author[3]{Irene Li}
\author[4]{Hiroshi Maruyama}
\author[5]{Hiroki Kanezashi}
\author[2]{Raquel Pérez-Arnal}
\author[4]{Kunihiko Miyoshi}
\author[6]{Euma Ishii}
\author[1]{Keita Suzuki}
\author[7]{Sayaka Shiba}
\author[8]{Mariko Kurokawa}
\author[9]{Yuta Kanzawa}
\author[10]{Naomi Nakagawa}
\author[11]{Masatoshi Hanai}
\author[3]{Yixin Li}
\author[3]{Tianxiao Li}

\affil[1]{The University of Tokyo}
\affil[2]{Barcelona Supercomputing Center}
\affil[3]{Yale University}
\affil[4]{Preferred Networks, inc.}
\affil[5]{Tokyo Institute of Technology}
\affil[6]{Tokyo Medical and Dental University}
\affil[7]{ESSEC Business School}
\affil[8]{Tokyo Metropolitan Hospital}
\affil[9]{Janssen Pharmaceutical}
\affil[10]{Kantar}
\affil[11]{SUSTech}

\begin{document}
{\setstretch{.8}
\maketitle

\date{June 10, 2020}

\begin{abstract}
This paper aims at providing the summary of the Global Data Science Project (GDSC) for COVID-19. as on May 31 2020. COVID-19 has largely impacted on our societies through both direct and indirect effects transmitted by the policy measures to counter the spread of viruses. We quantitatively analysed the multifaceted impacts of the COVID-19 pandemic on our societies including people’s mobility, health, and social behaviour changes.
\bigbreak
People’s mobility has changed significantly due to the implementation of travel re- striction and quarantine measurements. Indeed, the physical distance has widened at international (cross-border), national and regional level. At international level, due to the travel restrictions,the number of international flights has plunged overall at around 88 percent during March.In particular, the number of flights connecting Eu- rope dropped drastically in mid of March after the United States announced travel restrictions to Europe and the EU and participating countries agreed to close borders, at 84 percent decline compared to March 10th. Similarly, we examined the impacts of quarantine measures in the major city: Tokyo (Japan), New York City (the United States), and Barcelona (Spain). Within all three cities, we found the significant decline in traffic volume. In Japan, we found that different demographic groups respond differently to the governments’ messages, and people’s behaviour is affected more by the mood of the society than the official declaration of the state of emergency. In New York, through the analysis of the traffic volume of bicycle sharing services, we found out that traffic volume has significantly decreased after the beginning of March. In Barcelona, through the analysis of the use of the public bike system and the amount of traffic load within the metropolitan area, we detected that mobility was only significantly altered once the harshest measures were implemented, hinting at a potential inefficiency of mild measures. Furthermore, as the lockdown went on, the mobility kept decreas- ing, indicating an increasing adherence as the understanding of the severity increases. Overall, the mobility restrictions at different levels, both international and national, reduced the physical distance dramatically during the period, while there are regional and country-level differences in reaction.
\bigbreak
We also identified the increased concern for mental health through the analysis of posts on social networking services such as Twitter and Instagram. Notably, in the beginning of April 2020, the number of post with \#depression on Instagram doubled, which might reflect the rise in mental health awareness among Instagram users.
Besides, we also identified the changes in a wide range of people’s social behaviors through the analysis of Instagram data and primary survey data. From the analysis on Instagram data, we found out the rise in specific keywords related to COVID-19. The number of posts with \#cough started increasing in mid of March, several days before the stay-at-home orders in some countries, totaling 15,522 posts during March, 313 percent increase from the previous month. Other hashtags related to user behavioral changes, such as \#mask, \#facemask, \#stayalive, also increased in mid of March. The number of \#mask posted during March were up 441 percent from that during Febru- ary, while \#facemask, 296 percent and \#stayalive, 143 percent. We further explored the changes in perception and resulting behavioural changes in Japan. We found out that people’s perception regarding COVID-19 varies depending on the number of infected cases in the local community. Also, people’s behavior has some indicative signal for the future spreading the disease. Through these analysis, we believe that our quantitative examination of the impact of COVID-19 will contribute to the multifaceted and fundamental evaluation of COVID-19 impacts and the policies taken by countries.
 
\end{abstract}
}
\newpage
\tableofcontents
\newpage

\section{Introduction}
\subsection{Overview of GDSP for COVID-19}
The GDSP (Global Data Science Project) for COVID-19 consists of an international team focusing on various societal aspects including mobility, health, economics, education, and online behavior. The team consists of volunteer data scientists from various countries including the United States, Japan, Spain, France, Lithuania and China.

\subsection{Aim of GDSP for COVID-19}
The purpose of the GDSP is to quantitatively measure the impacts of the COVID-19 pandemic on our societies in terms of people's mobility, health, and behaviour changes, and inform public and private decision-makers to make effective and appropriate policy decisions.

\subsection{Project Scope }
a.	Quantifying Physical Distancing\\
physical  distancing is key to avoid or slow down the spread of viruses. Each country has taken different policies and actions to restrict human mobility. In this project, we investigate how policies and actions affect human mobility in certain cities and countries. By referencing our analysis of policy and secondary impacts, we hope that decision makers can make effective and appropriate actions. Furthermore, by analyzing human mobility, we also aim to develop a physical  distancing risk index to monitor the risk on areas with high population densities and probability of contraction. 
\bigbreak 
\noindent b.	Behavioral Changes\\
Due to physical  distancing and lockdown policies, people have begun relying on video conferencing tools for meetings, lectures, and conversations among friends more frequently than usual. Children are especially affected by the quarantine since many must refrain from going to their classrooms and take classes online. By leveraging various data sources, we will analyze how daily behavior has been affected by this pandemic, and also compare behaviors among different countries and cities. We will also measure online e-commerce and consumer behavior by analyzing sites such as Amazon.
\bigbreak
\noindent c.	Emotion Analysis\\
For health, we have focused on emotion changes that people have experienced during this pandemic. Emotion changes have stemmed from various reasons such as unemployment, implementation of stay-at-home policies, fear of the virus, etc. We quantify emotion changes by using social media data, including Twitter and Instagram. Since the breakout of COVID-19, we have seen an increase in online discussions that use hashtags such as \#COVID-19 and \#depression. We believe it is vital to visualize and analyze the differences in people’s perceptions of COVID-19. We also hope to analyze overall responses to the pandemic by sentiment: sadness, depression, isolation, happiness, etc. A further detailed analysis will also look into specific keywords and corresponding trends.

\subsection{Report Structure}
Each section in this report follows the following format; key takeaway, data description, policy changes, overall analysis, subcategory analysis.
\bigbreak
\bigbreak
\section{Quantifying Physical Distance}
We aim at analysing the seeming trade-off between economics and prevention of infection spread. Based on the calculation of physical distance index (mobility index), economic damages, and the number of newly infected patients, we evaluate the optimal level where we embrace both the steady decline in the number of infections and recovery of economics.

\subsection{Global Mobility Changes}
\subsubsection{Key Takeaway}
To investigate the effects of these travel restrictions on worldwide flights, we analyzed the decline of flights for continents and countries from public flight data. We found that the overall international flights significantly decreased from the beginning of March, at around 24 percent. In particular, the number of flights connecting Europe drastically dropped in mid of March after the United States announced travel restrictions to Europe and the EU and participating countries agreed to close borders, at 84 percent decline compared to March 10th. We conducted a more detailed analysis in another paper \cite{suzumura2020impact}.

\subsubsection{Data Description/ Methodology}
In order to analyze real-time international flight data, we obtained voluntarily provided flight dataset from The OpenSky Network - Free ADS-B and Mode S data for Research \cite{6846743},\cite{Martin}. The dataset contains flight records with departure and arrival times, airport codes (origin and destination), and aircraft types. The dataset includes the following flight information during January 1st to April 30th. The dataset for a particular month is made available during the beginning of the following month. The data covers 148 countries out of 195 countries, including 616 major airports and several small to medium size airports (Figure 1). The data was collected over the period of January 1 to March 31.
\bigbreak
As for a data methodology, we build a temporal network where a country or an airport is represented as a vertex and a connection between 2 countries or 2 airports is represented as an edge.  By building such a temporal network and compute shortest paths and their length between 2 countries or 2 airports, we can measure how travels are restricted in a quantitative manner by using graph analytics. The data is analysed on (1) global, (2) continental, (3) country (airport) level. The top 20 airports were based on the preliminary world airport traffic rankings released by ACI World \cite{Airport:}. 

\begin{figure}[H]
    \centering
        \includegraphics[scale=.2]{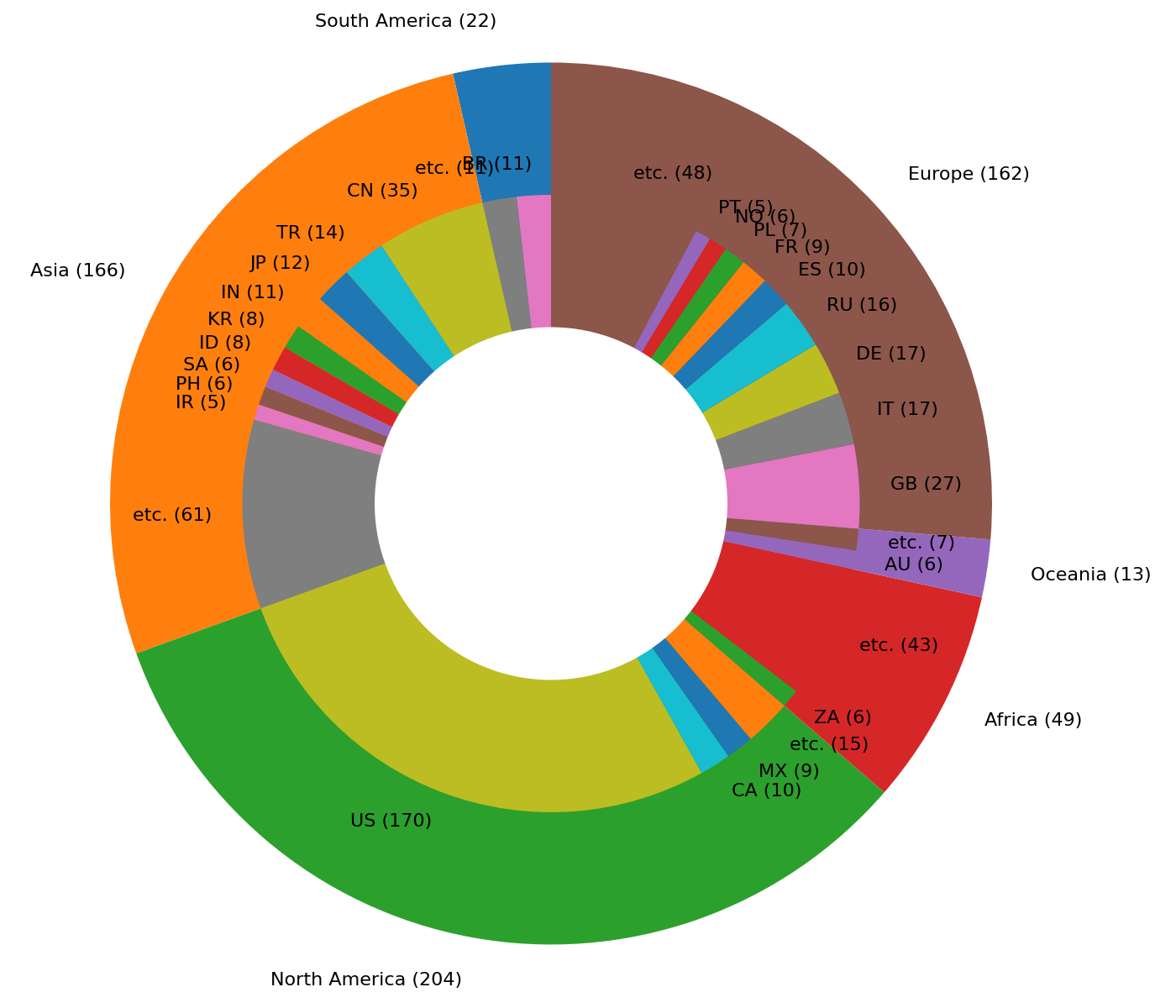}
    \caption{The Composition of Airports Examined}
    \label{fig:1}
\end{figure}

\subsubsection{Overall Analysis}
Before the end of February 2020, the overall number of (departed) international flights were around 8,000. However, since March 1 the number of international flights has started to decline. It reflected the first coronavirus death in the United States and the announcement of travel restrictions of ‘do not travel’ on February 29th. Since March 11, this decline has further accelerated in response to President Trump’s announcement of the travel restriction on 26 European countries on March 11 and 13.

\begin{figure}[H]
    \centering
        \includegraphics[scale=.13]{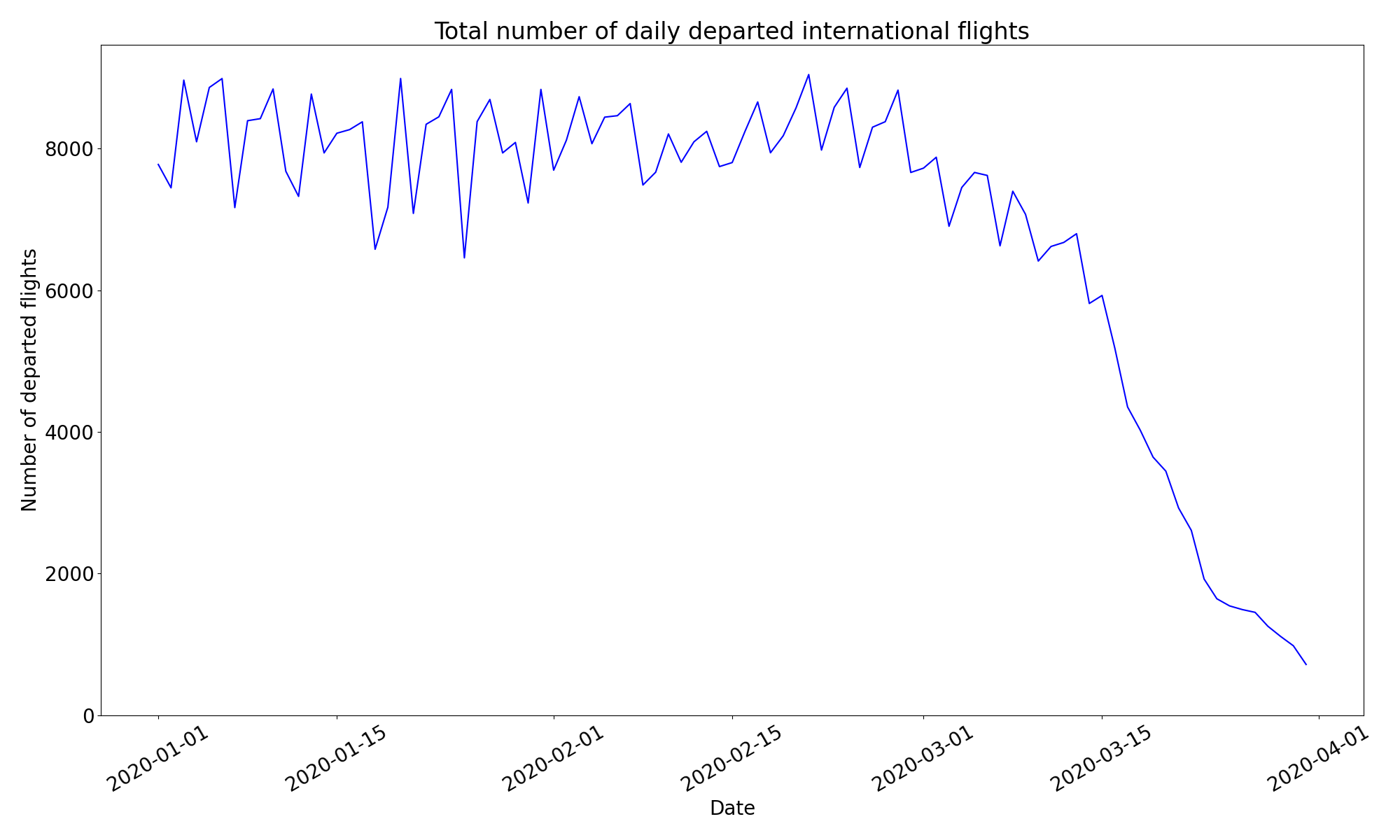}
    \caption{Total Number of Daily Departed International Flights}
    \label{fig:2}
\end{figure}

\subsubsection{Continental-level Analysis}
There are several turning points for declines of flights in some continents. On January 20th, the first cases outside mainland China occurred in Japan, South Korea and Thailand, according to the WHO, resulting in sudden decline of flights around January 20th in Asia. On February 23rd, Italy saw a major surge in coronavirus cases (up to 150), resulting in a decline of international flights in North America and Europe. On March 11th, WTO declared the COVID-19 as pandemic, and then international flights in all continents drastically decreased in the second half of March. 
\begin{figure}[H]
    \centering
        \includegraphics[scale=.2]{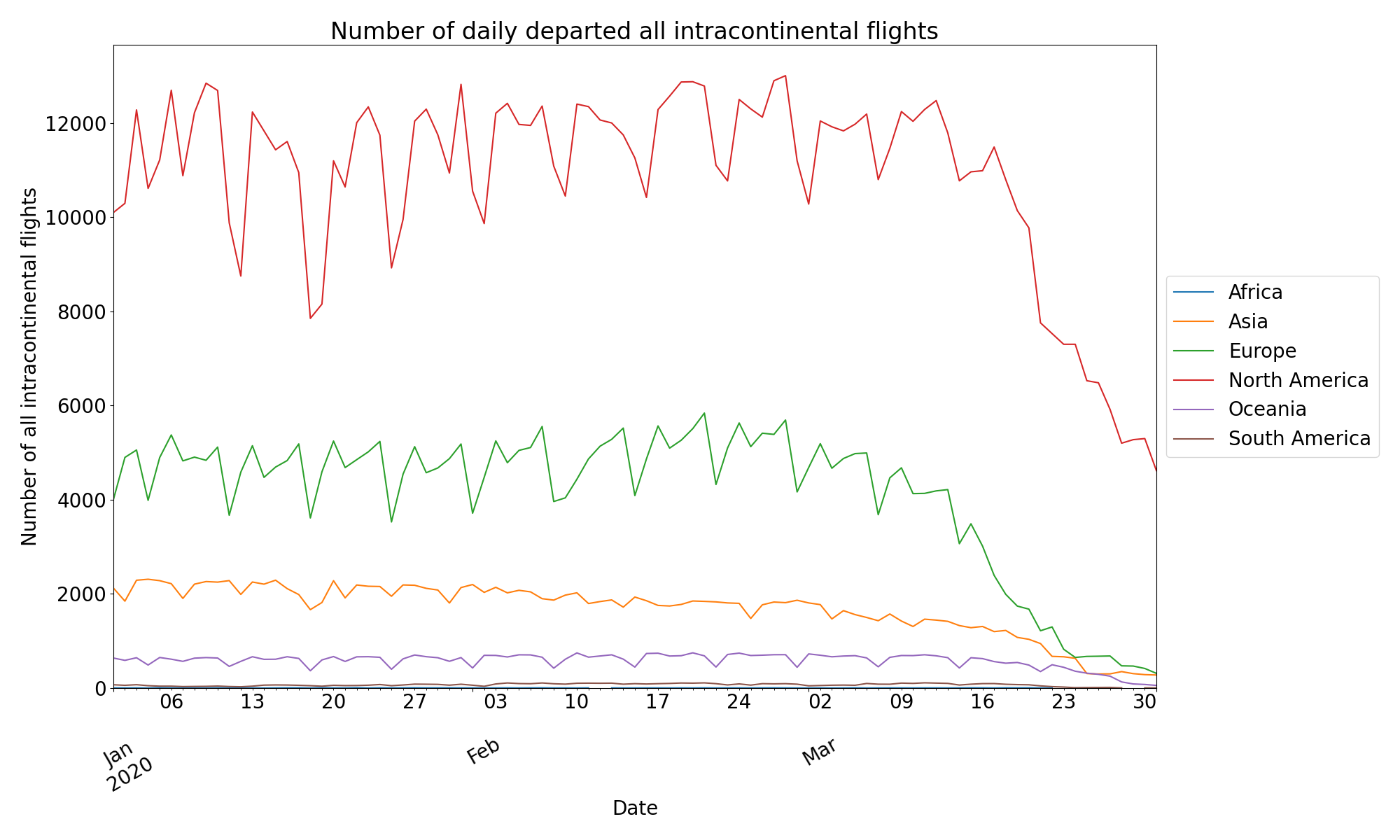}
    \caption{All intracontinental flights (including domestic flights)}
    \label{fig:3}
\end{figure}

\begin{figure}[H]
    \centering
        \includegraphics[scale=.2]{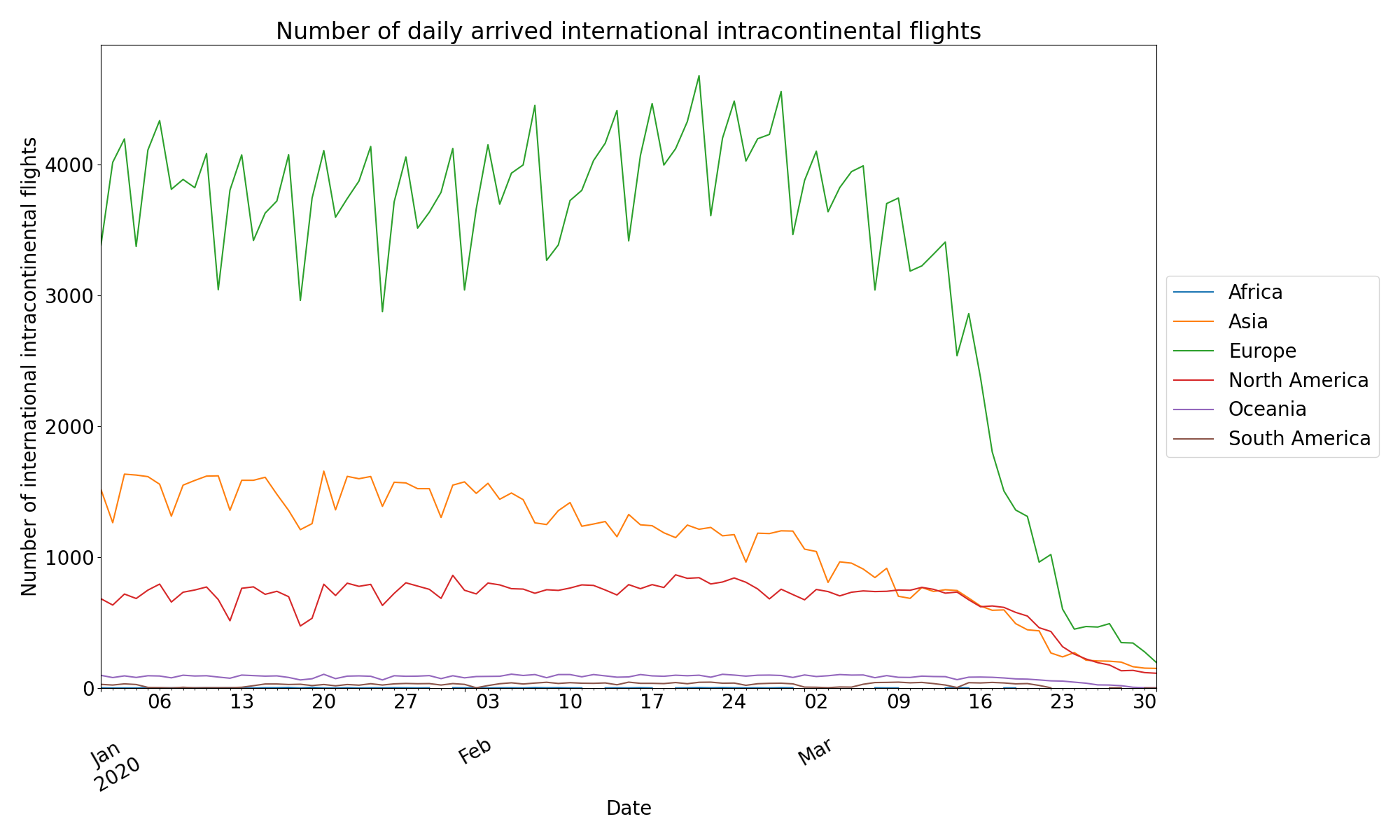}
    \caption{International intracontinental flights}
    \label{fig:4}
\end{figure}

\subsubsection{Country-level Analysis}
\begin{figure}[H]
    \centering
        \includegraphics[scale=.23]{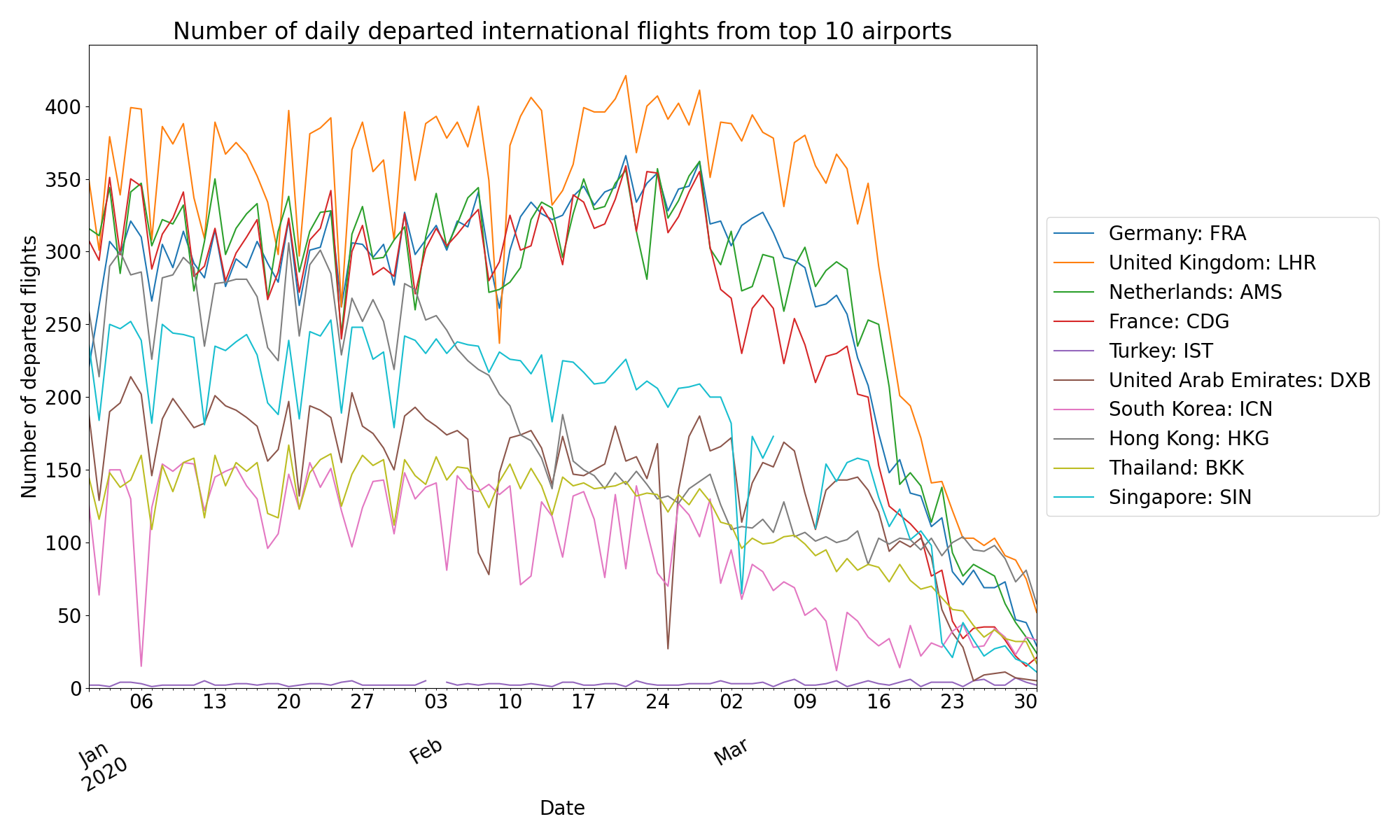}
    \caption{The absolute number of daily flights }
    \label{fig:5}
\end{figure}

\begin{figure}[H]
    \centering
        \includegraphics[scale=.2]{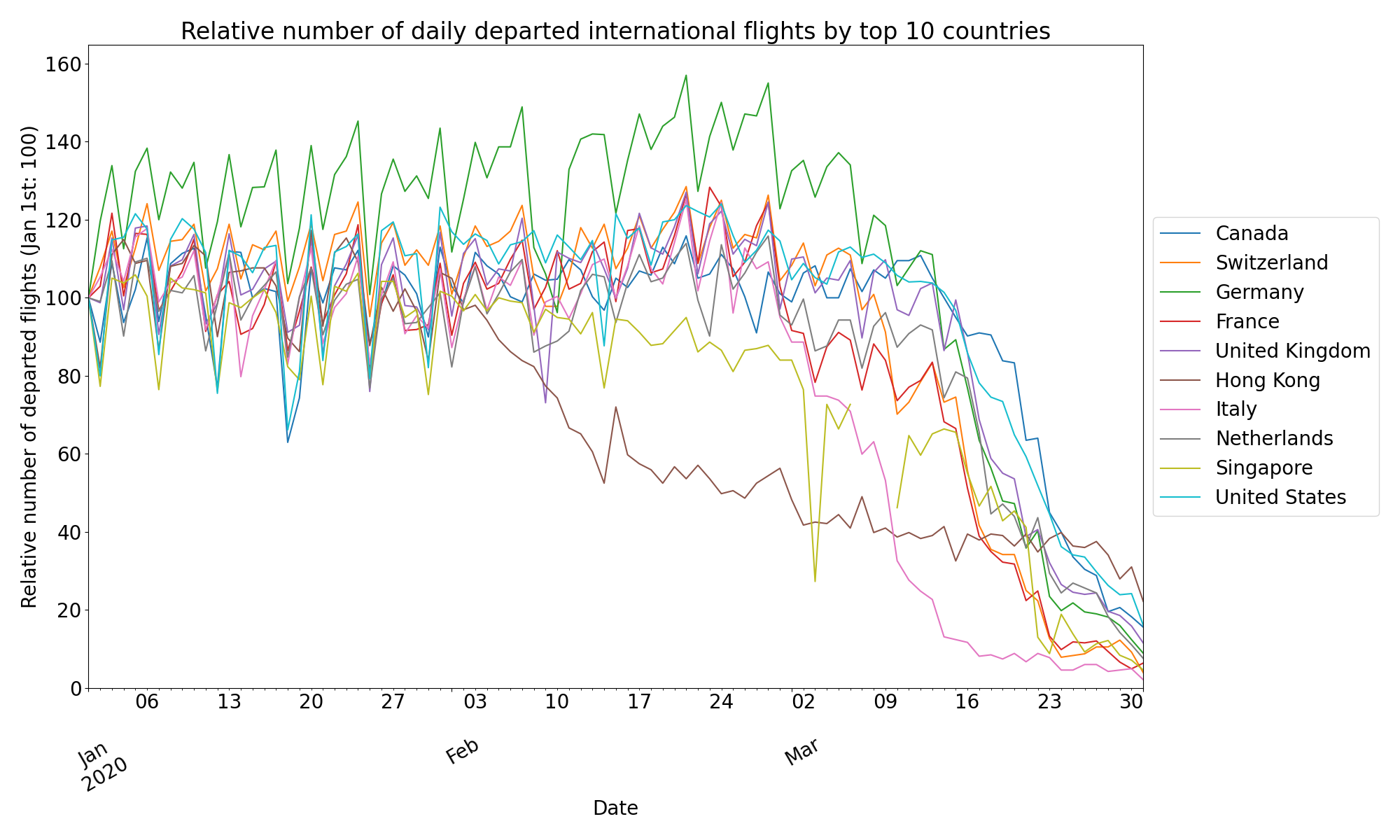}
    \caption{The relative number of daily flights (January 1st: 100)}
    \label{fig:6}
\end{figure}

Figure \ref{fig:5} and \ref{fig:6} describe the absolute and relative number of daily flights in top-10 countries with many international departures. From February 1st, the number of flights from Hong Kong gradually decreased after an announcement of tightening of border between China and Hong Kong was issued. The number of flights from Singapore suddenly dropped due to the confinement of the first case.
\bigbreak
In Italy, 28,800 people were infected before February 28th in Italy and cases in 14 other European countries remain an area of concern, and the US records its first coronavirus death and announces travel restrictions of ‘do not travel’ on February 29th, resulting in the decline of international flights from France Switzerland, and Italy around February 27th. Then, a significant slump of flights also occurred from around March 12th in these countries after the declaration of COVID-19 by WHO.

\begin{figure}[H]
    \centering
        \includegraphics[scale=.2]{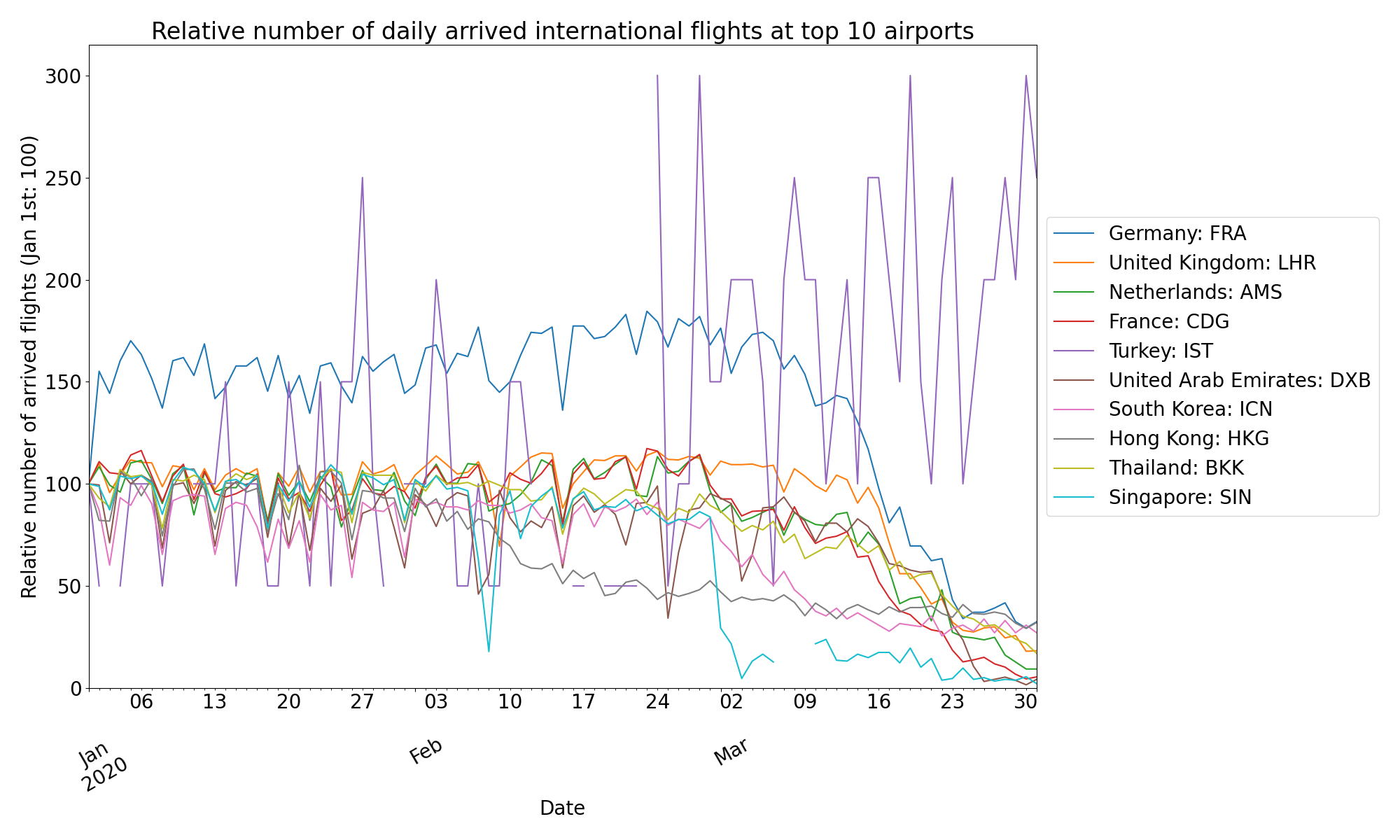}
    \caption{The absolute number of daily flights}
    \label{fig:7}
\end{figure}

\begin{figure}[H]
    \centering
        \includegraphics[scale=.2]{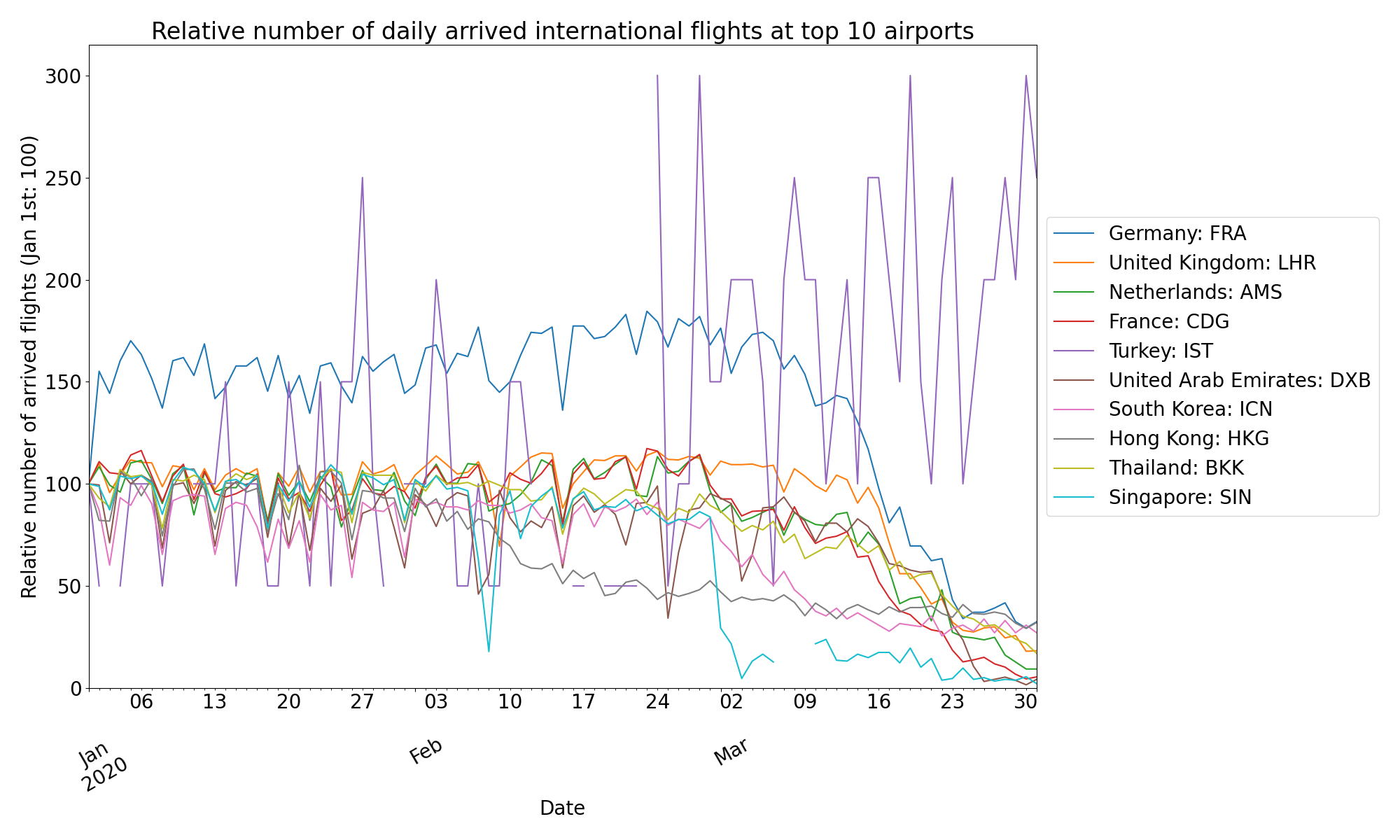}
    \caption{The relative number of daily flights (January 1st: 100) }
    \label{fig:8}
\end{figure}

Figure \ref{fig:7}and \ref{fig:8} describe the absolute and relative number of daily flights in top-10 airports with many international departures. The trend of declines is similar to the top-10 countries.

\begin{figure}[H]
    \centering
        \includegraphics[scale=.2]{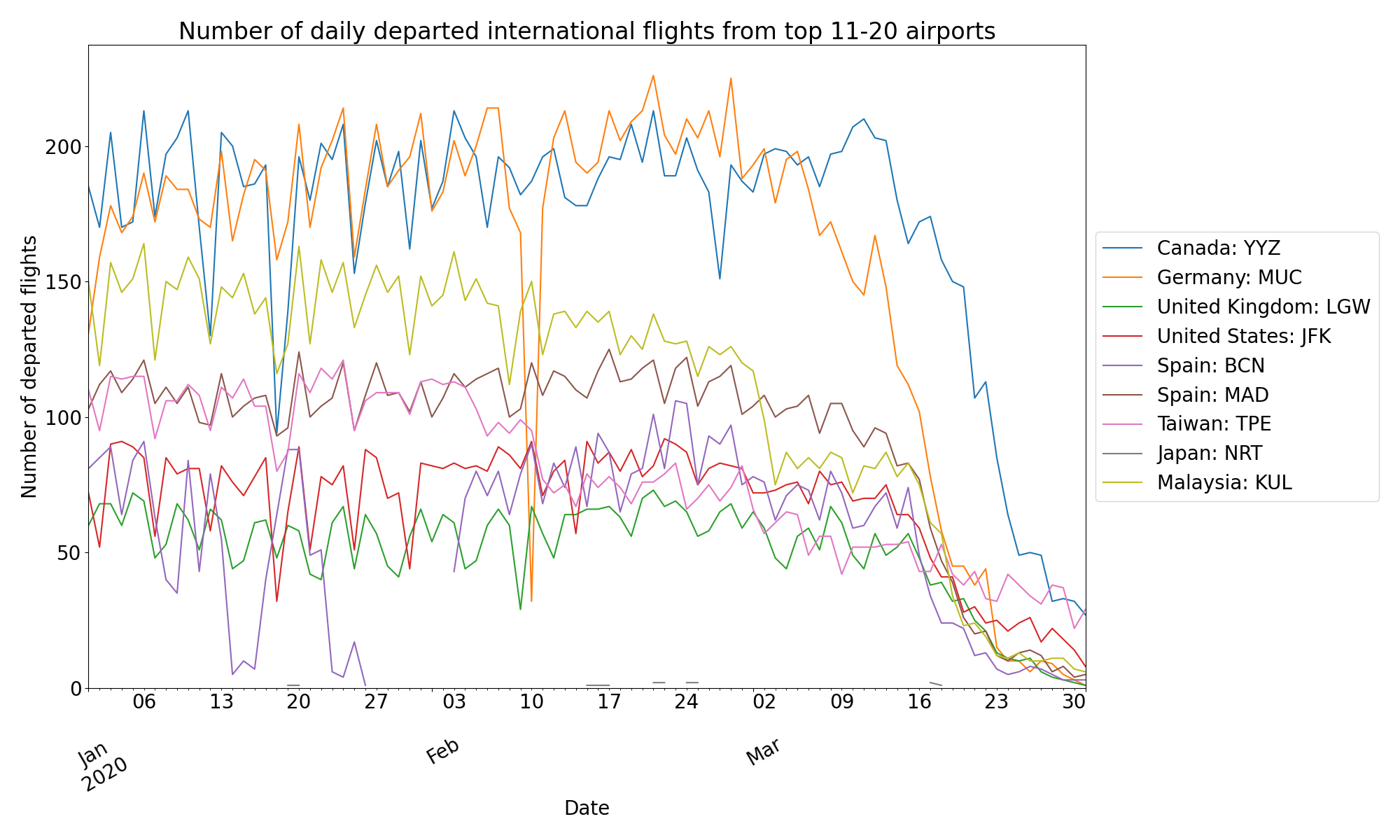}
    \caption{The absolute number of daily flights of Top 11-20 airports}
    \label{fig:9}
\end{figure}

\begin{figure}[H]
    \centering
        \includegraphics[scale=.2]{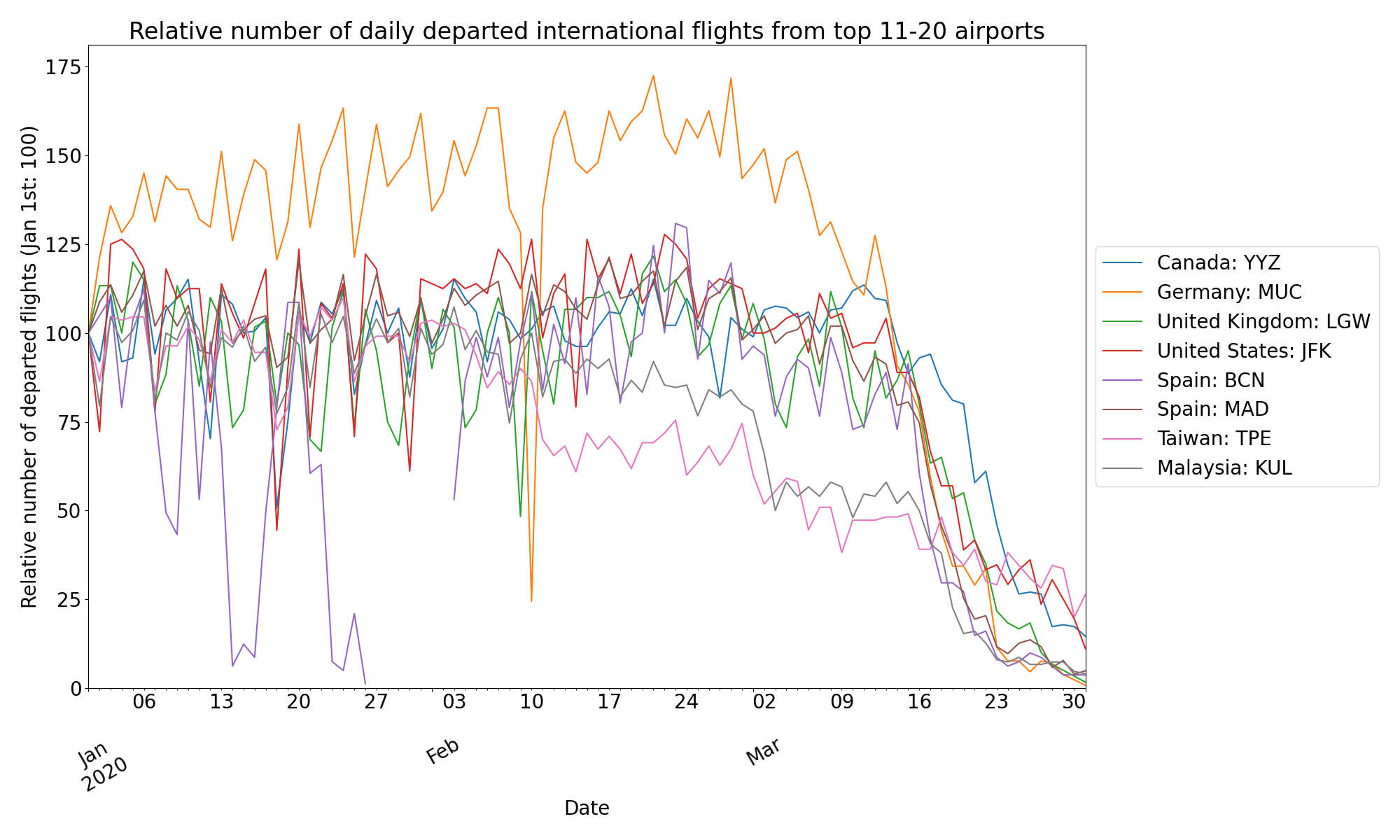}
    \caption{The relative number of daily flights of Top 11-20 airports (January 1st: 100) }
    \label{fig:10}
\end{figure}

Figure \ref{fig:9} and \ref{fig:10} describe the absolute and relative number of daily flights in top 11-20 airports with many international departures. The same slump of flights occurred as the top-10 airports and countries. In Taiwan, the number of departing flights gradually began to decline from February after WHO declared  a global health emergency and thousands of new cases in China on January 30th.

\subsection{Mobility Changes in Tokyo }
\subsubsection{Key Takeaway}
We examined the mobility in Tokyo and its relations to the national and local government’s measures to suppress COVID-19. We found that 1) different demographic groups respond differently to the governments’ messages (e.g., the senior responded first but the younger generations are willing to comply with the governments’ instructions to stay home once the formal announcement of the state of emergency is declared), and 2) people’s behaviour is affected more by the mood of the society than the official declaration of the state of emergency. We also investigated the correlation between the daily mobility index and the growth rate of reported confirmed cases, which suggested that the mobility index may be an early indicator of the growth rate of confirmed cases as well as the number of confirmed cases affecting future mobility. 

\subsubsection{Data Description/ Methodology}
We analyzed NTT DoCoMo data and accessed high-resolution hourly population data within Tokyo from Mobile Kukan Toukei \cite{docomo}. The data is based on mobile phone location on every hour and covers all the Tokyo metropolis (average daily population is around 11M) and from Jan. 1st, 2020 to the current date. We also received the same data for Jan. to Mar. in 2019 for the comparison.
\bigbreak
The data set divides Tokyo into 8,500 grid cells of 0.5km x 0.5km. The provided data is a collection of population vectors Pt where Pt[i] is the population of the grid cell i at time t (t is an hourly time point between 0:00 on January 1st and 23:00 on March 31st). We defined the overall mobility within Tokyo at time t as L1(Pt - Pt+1) where L1 is the L1 norm. Intuitively, this metric counts the sum of the number of people who came into or left each cell during the given hour. Note that these metric underestimates actual mobility since incoming and outgoing people within an hour cancel each other out. 
\bigbreak
The mobility index above for Tokyo includes large rural areas that may have contributed less to COVID-19 transmission. For this study, we clustered ~8,500 grid cells into 6 groups based on hour-of-day population patterns (each grid cell is represented as a 24-variable vector) using the 2019 data.

\subsubsection{Overall Analysis (City-level)}
\begin{figure}[H]
    \centering
        \includegraphics[scale=1]{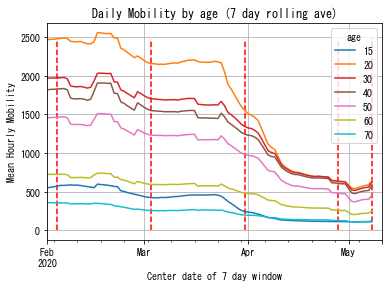}
    \caption{Daily mobility index of Tokyo by Age Groups}
    \label{fig:11}
\end{figure}

\begin{figure}[H]
    \centering
        \includegraphics[scale=1]{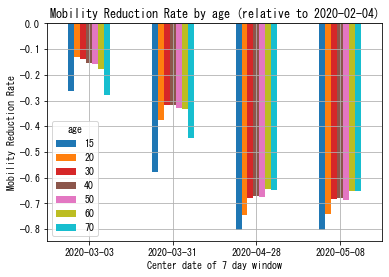}
    \caption{Mobility reduction rates by Age Groups}
    \label{fig:12}
\end{figure}

Figure \ref{fig:11} and \ref{fig:12} show the change of mobility index by age groups. We observe the largest drop of mobility occurred around March 25th when the media started to discuss potential “capitol lockdown”. When the official state of emergency was declared on April 9th, the mobility had already dropped to less than half of that of the normal time. We also note that the senior groups, who are supposed to be at higher risk, responded to the epidemic initially, but later the younger groups were more willing to stay at home.

\begin{figure}[H]
    \centering
        \includegraphics[scale=.67]{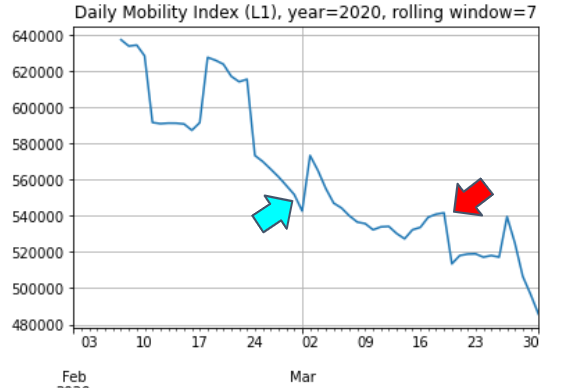}
    \caption{Daily mobility index of Tokyo in February and March 2020}
    \label{fig:13}
\end{figure}

\begin{figure}[H]
    \centering
        \includegraphics[scale=.7]{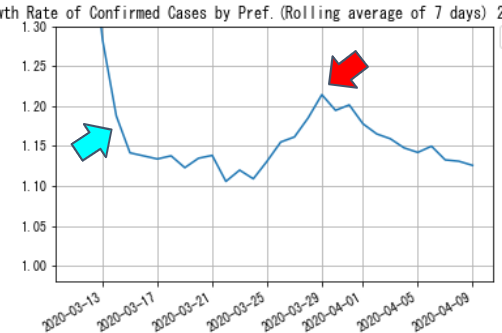}
    \caption{The daily growth rate of reported confirmed cases in Tokyo Metropolis}
    \label{fig:14}
\end{figure}

We also investigated the potential use of the mobility index as an earlier indicator of the future spread of the disease. Figure \ref{fig:13} and \ref{fig:14} shows the daily mobility index and the growth rate of confirmed cases in Tokyo.  In the plots, we noticed that the drop of mobility around March 2nd may be correlated to the drop of growth rate on March 14th as shown in the blue arrows, and the pickup of mobility on March 18th may be correlated to the peak of the growth rate on March 29th as shown in the red arrows It is not conclusive, but it may suggest that the mobility index has some signals for predicting the future spread of disease.

\subsubsection{Area-Level Analysis}
Different clusters show different responses in mobility to COVID-19. Figure 17 shows how mobility changes over time depending on the cluster.

\begin{figure}[H]
    \centering
        \includegraphics[scale=1]{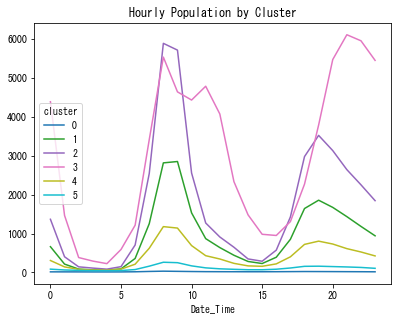}
    \caption{Daily mobility index of Tokyo in February and March 2020}
    \label{fig:15}
\end{figure}

\begin{figure}[H]
    \centering
        \includegraphics[scale=.3]{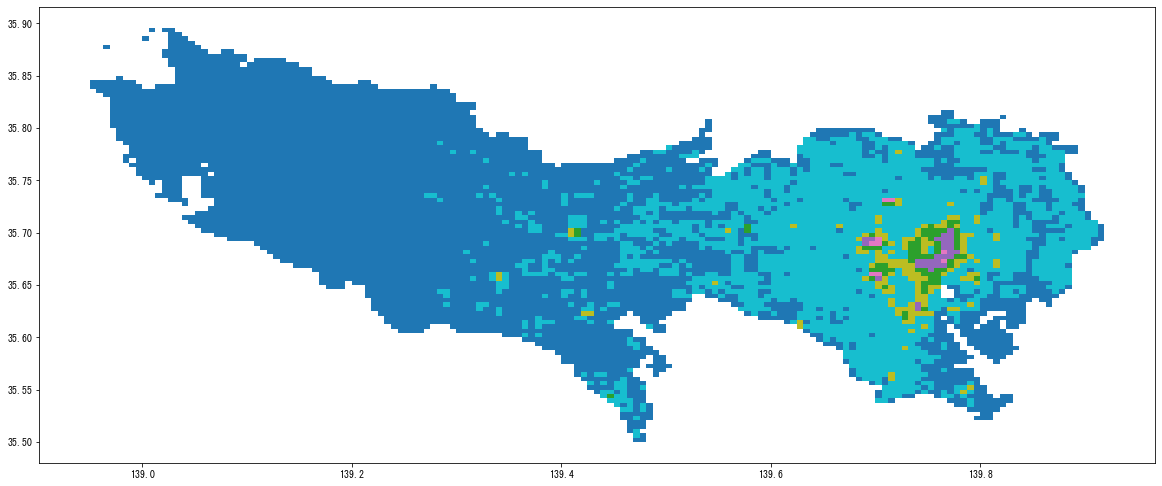}
    \caption{The daily growth rate of reported confirmed cases in Tokyo Metropolis}
    \label{fig:16}
\end{figure}

\begin{figure}[H]
    \centering
        \includegraphics[scale=1]{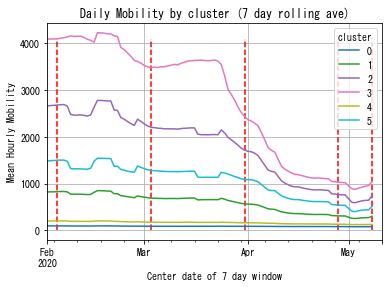}
    \caption{Daily Mobility by Cluster}
    \label{fig:17}
\end{figure}

\begin{figure}[H]
    \centering
        \includegraphics[scale=1]{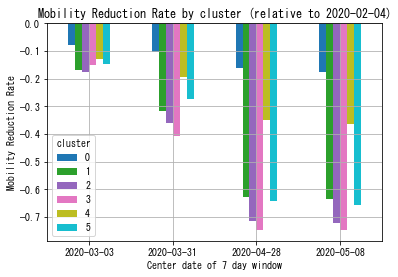}
    \caption{Mobility Reduction Rate by Cluster}
    \label{fig:18}
\end{figure}

\subsection{Mobility Changes in New York City }
\subsubsection{Key Takeaway}
We analyzed the changes in traffic volumes and a bicycle sharing service in New York City to examine the effect of COVID-19 and announcements from the city government. We found out the traffic volume has significantly decreased after the beginning of March, and thousands of people use CitiBike in every daytime.

\subsubsection{Data Description/ Methodology}
\begin{figure}[H]
    \centering
        \includegraphics[scale=.8]{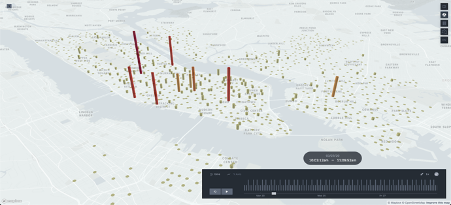}
    \caption{Interactive visualization dashboard of the number of used bikes}
    \label{fig:19}
\end{figure}

\begin{figure}[H]
    \centering
        \includegraphics[scale=.8]{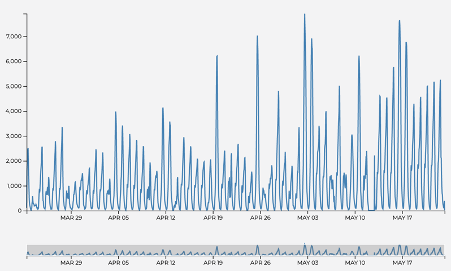}
    \caption{Total hourly number of departed bikes from all Citi Bike stations in NYC}
    \label{fig:20}
\end{figure}

We analyzed the mobility changes in New York City through the road traffic data and tracking data of sharing bikes.
We retrieved public historical data about traffic volume of freeways from NYC Open Data \cite{NYCTraffic}. As for road traffic data, we extracted the daily average travel time and speed in 20 days ranges. The x-axis of Figure \ref{fig:19} and Figure \ref{fig:20} shows the number of days from the first Sunday of March (Blue lines: 2019 3/3 - 3/21, Orange lines: 2020 3/1 - 3/19). We observed 3 representative locations within New York City: Lincoln Tunnel, Robert F. Kennedy Bridge, and Brooklyn Bridge. There overall was a significant decrease in traffic volumes from March 11th.
\bigbreak
As for sharing bike data, we track the number of people using bikes at each bike station every 30 seconds from NYC Open Data \cite{NYCBike}. We aggregated the number of departed bikes in 938 bike stations, located in most areas of Manhattan, Brooklyn and Queens near the East River, and Jersey City along the Hudson River. Since we can only track the number of available bikes in each station, we estimate the number of departed bikes by computing the difference in the number of available bikes between two timestamps. We developed an interactive visualization dashboard that illustrates how bikes are used over time since March 23th.

\subsubsection{Policy Changes }
After the NYC municipality recommended citizens to ride bikes instead of using public transportation, there was a surge in the usage of Citi Bike in the beginning of March, a privately-owned public bicycle sharing system in New York City \cite{Gersh}.

\subsubsection{Overall Analysis}
\begin{figure}[H]
    \centering
        \includegraphics[scale=.6]{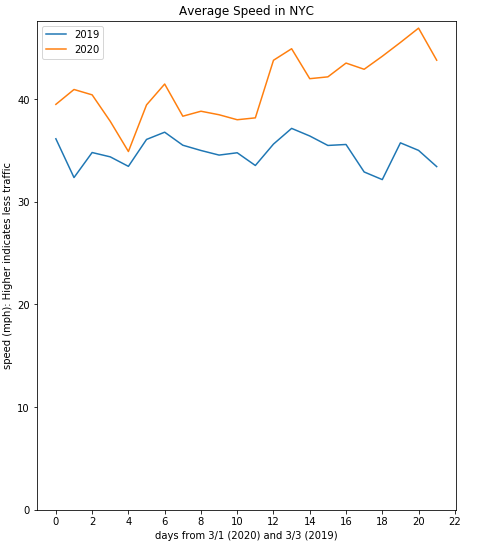}
    \caption{Average Speed and Travel Time in NYC}
    \label{fig:21}
\end{figure}

\begin{figure}[H]
    \centering
        \includegraphics[scale=.6]{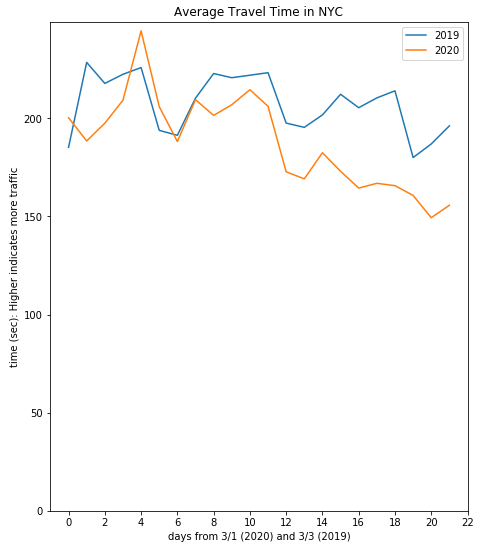}
    \caption{Average Speed and Travel Time in NYC}
    \label{fig:22}
\end{figure}

Figure 18 describes the total number of hourly CitiBike usages from March 23rd. More than a thousand of people used CitiBike in a peak hour every day, and in some days the number of peak usages exceeded 6,000 in some days (e.g., April 19th, 25th, and May 2nd). From the beginning of May, more than 4,000 people used bikes in peak hour.
\bigbreak
After the first confirmed case of COVID-19 in New York City on March 1st, freeway traffic drastically decreased. The overall daily average traffic speed in NYC in 2020 (Figure \ref{fig:21}) is higher than that of 2019 (31 percent faster on March 23rd), and the average travel time (Figure \ref{fig:22}) on the freeway became significantly shorter after March 13th in comparison to 2019 (20 percent shorter on March 23rd). Similar trends are found in Manhattan, the Bronx, and Queens. In Brooklyn, the roads gradually became vacant after March 5th (20 percent higher speed and 18 percent shorter travel time on March 23rd). In Staten Island, the traffic speed is higher (64 percent on March 23rd) but the travel time has become overall longer (58 percent on March 23rd) than that of 2019.

\subsection{Mobility Changes in Barcelona }
\subsubsection{Key Takeaway}
We analyzed the use of the public bike system and the amount of traffic load within the Barcelona metropolitan area, to understand how the CoVid19 pandemic and the government measures were affecting public movement. We detected that mobility was only significantly altered once the harshest measures were implemented, hinting at a potential inefficiency of mild measures. Furthermore, as the lockdown went on, the mobility kept decreasing, indicating an increasing adherence as the understanding of the severity increases. 
\bigbreak
The mobility data in Barcelona, Spain was collected through the location signal of FaceBook application users who have consented to share their location, public bike sharing data, and road traffic data. Similar to New York City, public bike sharing is available in Barcelona. By analyzing the availability of docking stations throughout the city, we measure the changes in population mobility. Traffic data is obtained from the open data released by the Barcelona city hall. It includes over 100 measuring points, evenly distributed throughout the city.

\subsubsection{Data Description/ Methodology}
The mobility data in Barcelona, Spain was collected through the location signal of FaceBook application users who have consented to share their location, public bike sharing data, and road traffic data. Similar to New York City, public bike sharing is available in Barcelona. By analyzing the availability of docking stations throughout the city, we measure the changes in population mobility. Traffic data is obtained from the open data released by the Barcelona city hall. It includes over 100 measuring points, evenly distributed throughout the city.

\subsubsection{Policy Changes}
The first detected case of COVID-19 within Spain was on January 31st in the Canary Islands, located more than 1,000 km from peninsular Spain. By late February, imported cases were detected in the mainland, and on February 26th the first endemic case was diagnosed. On March 9th, 999 cases were diagnosed and certain regions in Spain started implementing local restriction policies. By March 13th, cases had been detected across all 50 provinces. The following day, March 14th, the Spanish Government announced a state of emergency, and implemented a lockdown for the whole population. Citizens were only permitted to travel for work, and all social events were prohibited. This lockdown was reinforced on March 29th with total mobility restrictions, and only essential services were an exception.
\bigbreak
The first restriction against COVID-19 by the regional government of Barcelona was on March 11th that informed citizens to avoid gatherings of more than 1,000 people. Two days later, on March 13th, with 508 confirmed cases in the region, all classes were suspended. By March 14th, a national lockdown was declared by the Spanish government.  The effects on mobility are only visible from March 13th, indicating that the local population did not alter their mobility patterns in response to earlier and milder governmental restrictions.

\subsubsection{Overall Analysis}
(a) Mobility Analysis based on Facebook App User Data 
The mobility was the same as previous years until the week of March 9th. However, mobility decreased to 34 percent on average after the national lockdown. This continued to decrease during the second week of lockdown, indicating a stronger adherence as the severity of the situation became more apparent. By the week of March 30th, mobility fluctuates between 15-20 percent compared to pre-COVID-19 mobility.

\begin{figure}[H]
    \centering
        \includegraphics[scale=.7]{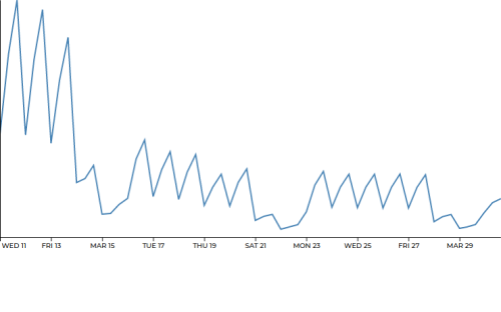}
    \caption{Mobility of Facebook App Users}
    \label{fig:23}
\end{figure}

(b) Mobility Analysis based on Public Bike Service in Barcelona
The first measures affecting Barcelona were implemented on March 13th, after which the availability of stations was 89 percent on average when compared to the previous week. On March 14th, availability was down to 62 percent in comparison to the previous week. March 15th saw minimal activity, and bike services were indefinitely suspended on March 16th.

\begin{figure}[H]
    \centering
        \includegraphics[scale=.24]{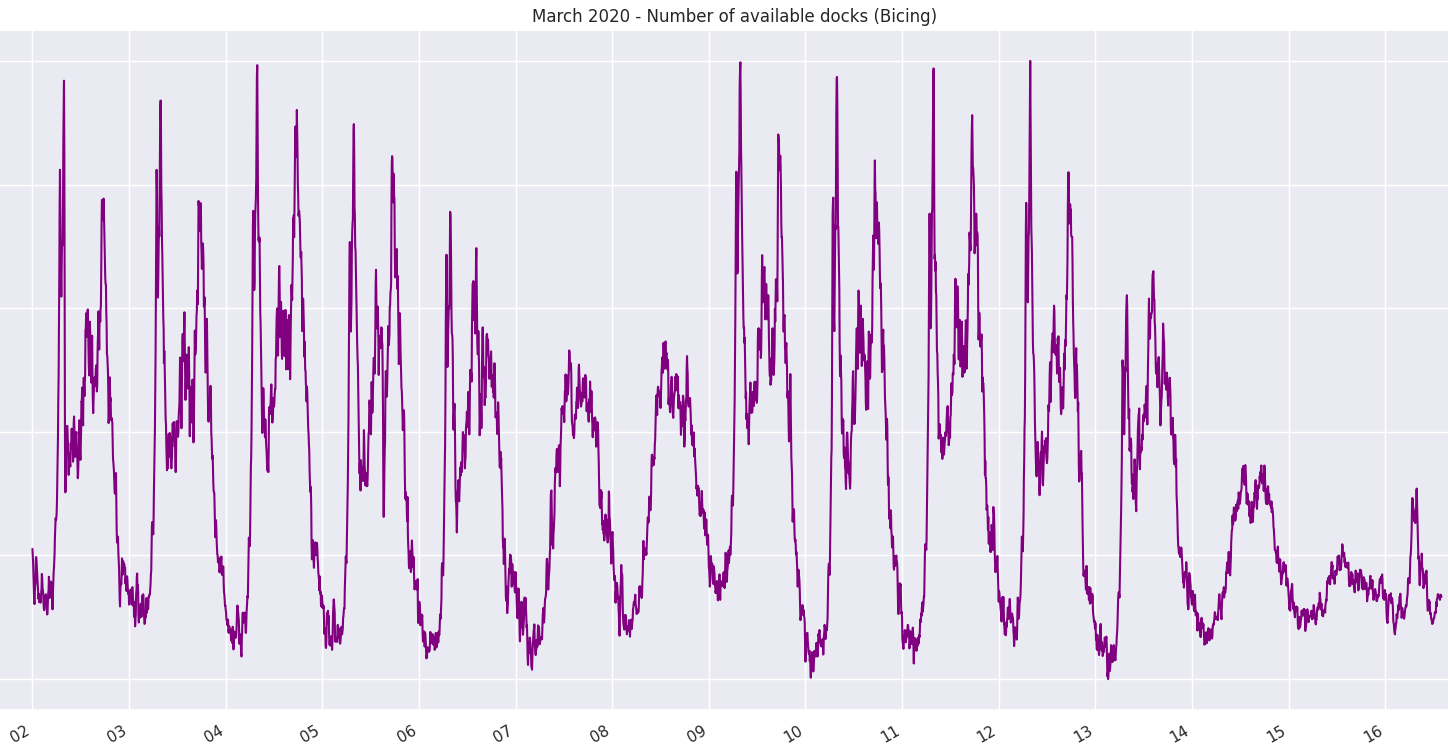}
    \caption{The Number of Available Docks (Bicing)}
     \footnote{Aggregated level of occupancy of bike stations, by day. Source:https://opendata-ajuntament.barcelona.cat/data/ca/dataset/estat-estacions-bicing}
    \label{fig:24}
\end{figure}

The road traffic has increased on the weekends, especially on Sundays (Figure 25).

\begin{figure}[H]
    \centering
        \includegraphics[scale=.2]{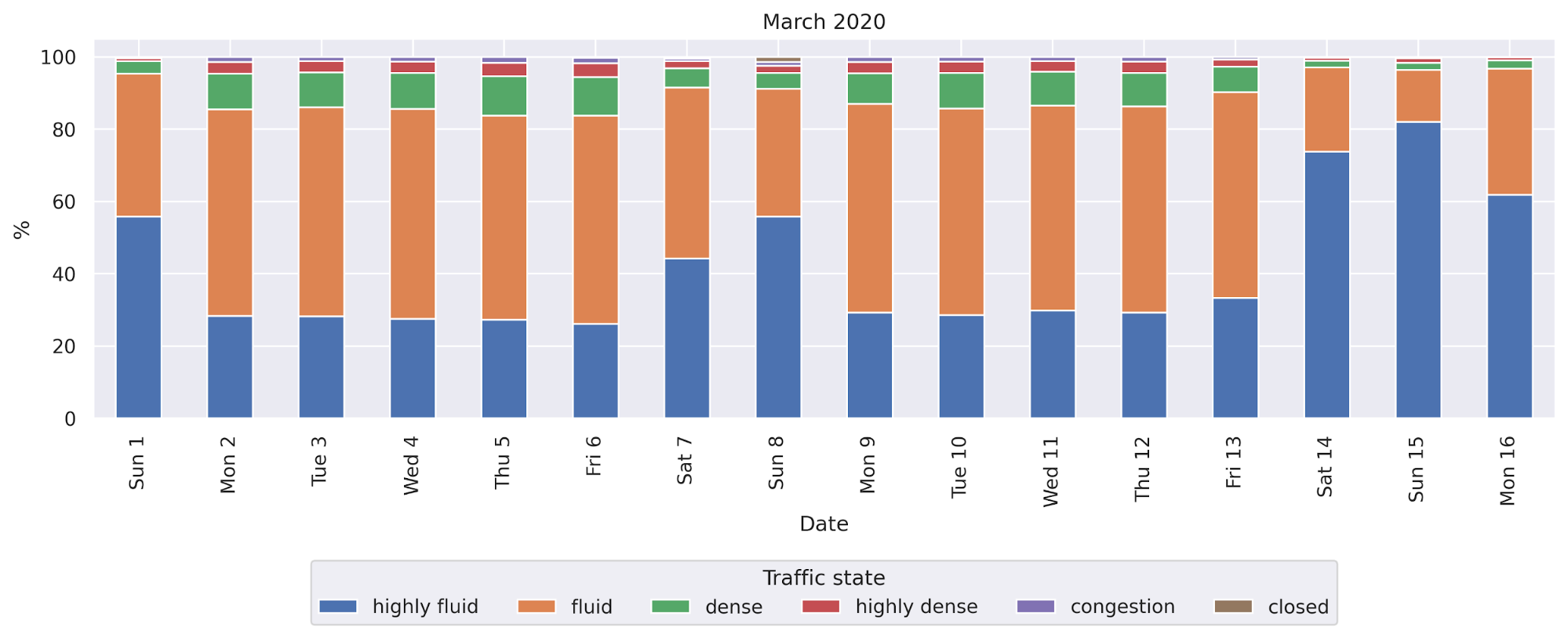}
    \caption{Road Traffic in Barcelona in March 2020}
    \footnote{Aggregated status of 100 road sectors in Barcelona, from March 2019. Source https://opendata-ajuntament.barcelona.cat/data/en/dataset/trams}
    \label{fig:25}
\end{figure}

The same data source showed minor changes in road congestion until March 13th, and then a dramatic change after the national lockdown was announced on March 14th (Figure 26). 

\begin{figure}[H]
    \centering
        \includegraphics[scale=.2]{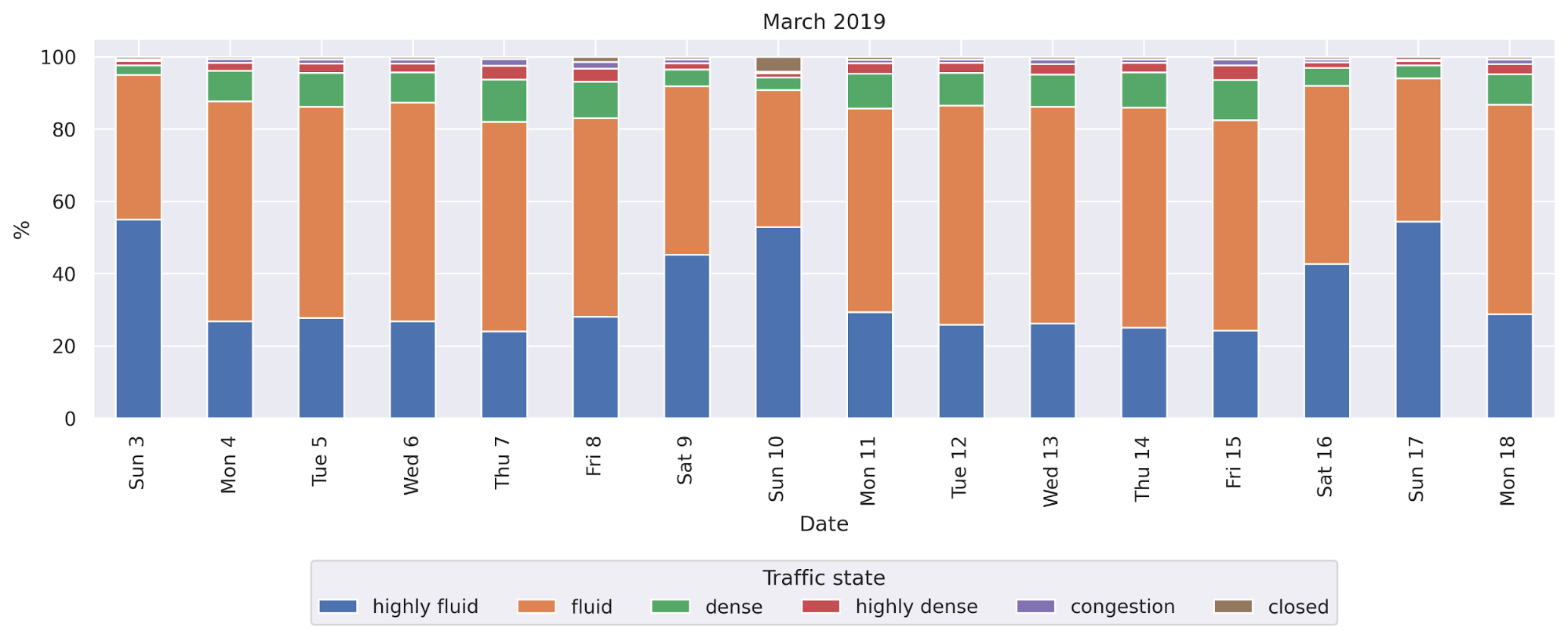}
    \caption{Road Traffic in Barcelona in March 2019}
    \footnote{Aggregated status of 100 road sectors in Barcelona, from March 2020, Source: https://opendata-ajuntament.barcelona.cat/data/en/dataset/trams}
    \label{fig:26}
\end{figure}

\subsection{Analysis of Physical Distance on Instagram}
\subsubsection{Key Takeaway}
Posts with some hashtags related to physical  distancing suddenly started increasing in mid of March 2020. A hashtag \#zoom, online meeting software, was frequently posted all over the world in late March, and the stock price of the software company rose dependent upon the increasing volume of posts with the hashtag.

\subsubsection{Data Description/ Methodology}
A post on Instagram would be summarized and categorized by hashtags on the post. We analyse the following hashtag, which represent physical  distancing have increased significantly since March 2020; \#stayhome, \#stayathome, \#socialdistancing, \#workfromhome, \#zoom (online meeting software).

\subsubsection{Overall Analysis }
As of April 11, 2020, more than 16 million posts with \#stayhome have uploaded on Instagram, including 6 million posts in March 2020. During mid of March 2020, the number of posts with \#stayhome gradually rose up, with states across the U.S. announcing a stay-at-home order. On March 24, 2020, Instagram announced that it launched a “Stay Home” sticker to help those practicing physical  distancing connect with others. This might have also boosted the number in March 2020.

\begin{figure}[H]
    \centering
        \includegraphics[scale=.8]{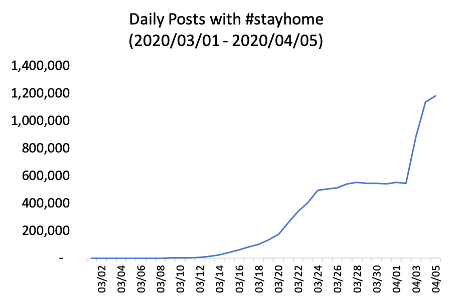}
    \caption{Daily Posts with \#stayhome}
    \label{fig:27}
\end{figure}

We have analyzed \#socialdistancing posted with positional information among 292,163 \#socialdistancing from April 3, 3 p.m. to April 5, 3 p.m. (UTC). Top countries tagged with the hashtag in this period are the US, the UK, Canada, Indonesia, India, Australia, and Germany. Posts with the hashtag were posted during local daytime, mainly in late afternoon in each country.

\begin{figure}[H]
    \centering
        \includegraphics[scale=.8]{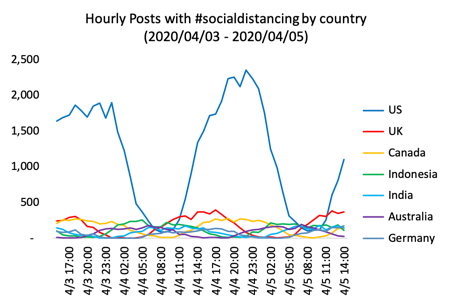}
    \caption{Hourly Posts with \#scialdistancing by country}
    \label{fig:28}
\end{figure}

\begin{figure}[H]
    \centering
        \includegraphics[scale=.8]{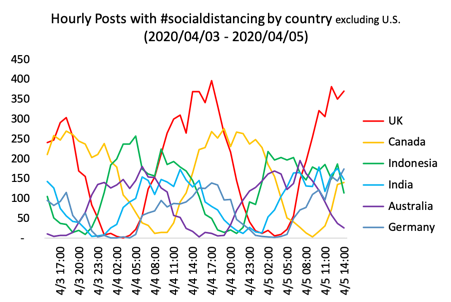}
    \caption{Hourly Posts with \#scialdistancing by country excluding the United States}
    \label{fig:29}
\end{figure}

Since some states and counties introduced work from home measures in response to the COVID 19 pandemic in mid of March 2020, the number of posts with \#zoom, online meeting software, has been increasing. The hashtag tends to increase on weekdays rather than weekends.  The stock price of Zoom Video Communications, Inc. (ZM) at the Nasdaq rose up along with the number of posts on Instagram until the end of March. These facts might suggest that people started to work on Zoom on weekdays, which increased the value of Zoom in the stock market by more than 50 percent.

\begin{figure}[H]
    \centering
        \includegraphics[scale=.8]{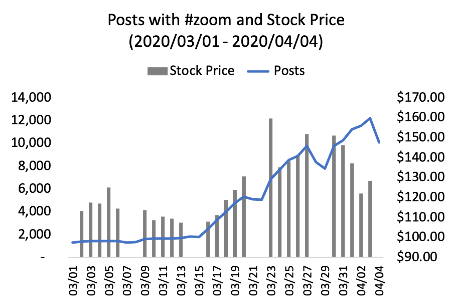}
    \caption{Posts with \#zoom and Stock Price}
    \label{fig:30}
\end{figure}

\bigbreak
\bigbreak

\section{Emotion Analysis}
\subsection{Emotion Analysis on Twitter}
\subsubsection{Key Takeaway}
The outbreak of COVID-19 recently has affected human life to a great extent. Besides direct physical and economic threats, the pandemic also indirectly impacts people’s emotional conditions, which can be overwhelming but difficult to measure. We apply natural language processing (NLP) \cite{vaswani2017attention}, \cite{qi2020stanza} to analyze tweets in terms of emotion analysis and attempt to find more in-depth topics and facts about emotions in terms of COVID-19 \cite{li2020depressed}.

\subsubsection{Data Description/ Methodology}
We have seen an increase in discussions tagged with hashtags such as \#COVID-19 and \#depression on Twitter and Instagram. We believe it is vital to analyze differences in. We also hope to analyze overall responses to the pandemic as well as changes in behavior due to the virus and generate reports on global situations regarding mental health. We plan to categorize the tweets and Instagram posts that mention COVID-19 by sentimental categories: .anger, anticipation, disgust, fear, joy, sadness, surprise and trust. Further detailed analysis will also look into specific keywords and corresponding trends.
\bigbreak
We  applied  Twitter  API to  conduct  a  crawler  with  a  list  of  keywords: \#coronavirus,  \#covid19,  \#covid, \#COVID-19,  \#covid19,  \#confinamiento,  \#flu,  \#virus,  \#hantavirus,  \#fever,  \#cough,  \#social  \#distance,  \#lockdown, \#pandemic, \#epidemic, \#conlabelious, \#infection, \#stayhome, \#corona, \#epidemie, \#epidemia,\begin{CJK*}{UTF8}{gbsn}新冠肺炎\end{CJK*},
 \begin{CJK}{UTF8}{gbsn}新型冠狀病毒\end{CJK},
 \begin{CJK}{UTF8}{gbsn}疫情\end{CJK},
\begin{CJK}{UTF8}{gbsn}新冠病毒\end{CJK}, 
\begin{CJK}{UTF8}{gbsn}感染\end{CJK},
\begin{CJK}{UTF8}{min}新型コロナウイルス\end{CJK},
\begin{CJK}{UTF8}{min}コロナ\end{CJK}. \\ Each day, we are able to crawl 3 million tweets in free text format from different languages. Due to the high capacity, we look at the tweets from March 24 to 26, 2020 to get language and geolocation statistics.  Among these tweets, 8,148,202 tweets have the language information (“lang” field of the “Tweet” Object in Tweet API), and 76,460 tweets have the geographic information (“country\_code”value from the “place” field if not none).
\bigbreak
A. \textbf{Sentiment Distribution on 8 Categories}:\\
We applied a deep learning model (BERT) \cite{devlin2018bert} trained on 750 manually labeled cases to 1 million English tweets. Fear, Anger and Sadness ranked first.
\bigbreak
B. \textbf{Sentiment Trend among Topics}:\\
We now look at the emotion trends on different topics. Using BERT, we analyzed two topics: “mask” and “lockdown”. 
\bigbreak
C. \textbf{The reason why people feel sad or fear}:\\
To understand why people feel fear and sadness, we calculated correlation on the tweets categorized by fear and sadness, and then kept nouns and noun phrases with the help of Stanford Stanza tool.
\bigbreak
D. \textbf{The topics of Tweets}:\\
We utilized the LDA(Latent Dirichlet allocation)  topic modeling \cite{10.5555/944919.944937} to analyze the topics on people’s tweets. Each “topic” learned by the model is a bunch of key words, then we manually labeled these topics as meaningful concepts. 

\subsubsection{Overall Analysis}
\begin{figure}[H]
    \centering
        \includegraphics[scale=.4]{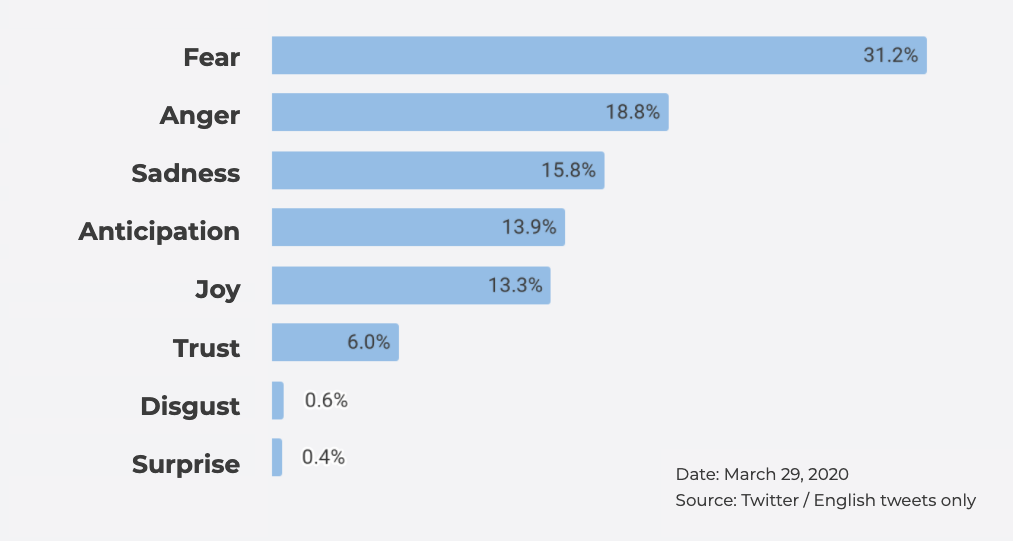}
    \caption{Emotion Distribution }
    \label{fig:31}
\end{figure}

A. \textbf{Sentiment Distribution on 8 Categories}:\\
By applying our model, we show the emotion distribution among 8 categories in Fig. \ref{fig:31}. Each day, the overall distribution has no big difference. So we show the results on a 1 million tweets from March 29th, 2020. Note that these tweets contain many languages rather than English. We could notice that the top emotions are very positive: fear, anger and sadness. 

\begin{figure}[H]
    \centering
        \includegraphics[scale=.4]{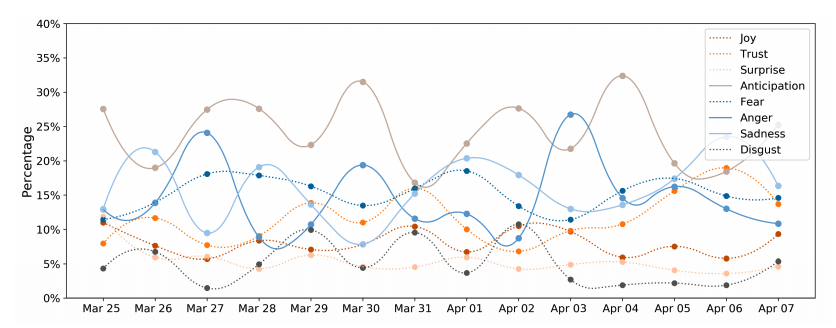}
    \caption{Emotion trend over the time about "mask"}
    \label{fig:32}
\end{figure}

\begin{figure}[H]
    \centering
        \includegraphics[scale=.4]{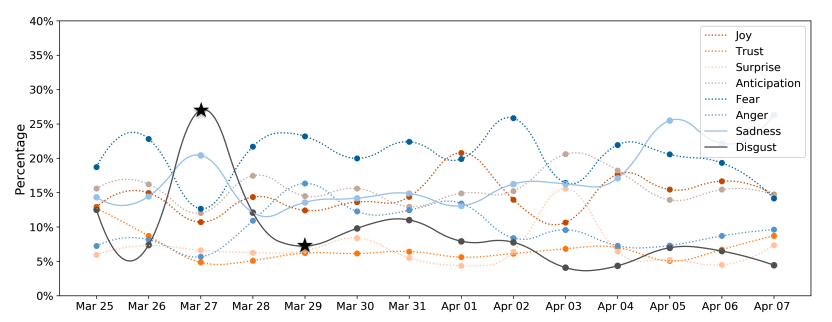}
    \caption{Emotion Trend over the Time about “lockdown”}
    \label{fig:33}
\end{figure}

\noindent B. \textbf{Sentiment Trend among Topics}:\\
We select the data of two weeks (March 25, 2020-April 7, 2020), and apply our model to predict the emotions on all the tweets we crawled (around 3 million each day) that contain the two “masks” and “lockdown” respectively. We found the dominating emotions and variations of the change are closely related to the topic. In Fig. \ref{fig:32} and \ref{fig:33}, we illustrate the emotion trend for each single day of the selected keywords. The high variation (plot in solid lines in the figures) showed up in sadness, anger and anticipation for the tweets that contain the word “mask” in Fig.\ref{fig:32}, and disgust, sadness for the tweets that contain the word lockdown in Fig. \ref{fig:33}. Especially for the lockdown tweets, the percentage of disgust emotion had a significant increase on March 27 and dropped on the next two days, as marked with the black asterisks. 
\bigbreak
To further investigate, we looked at the news on March 27, which included the U.S. as the first country to report 100,000 confirmed coronavirus cases, and 9 in 10 Americans were staying home; India and South Africa joined the countries to impose lockdowns. Given that the United States, India and Brazil have large groups of twitter users, we assume that this dramatic change may be triggered by that news.

\begin{figure}[H]
    \centering
        \includegraphics[scale=.8]{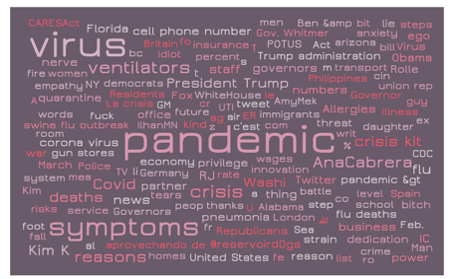}
    \caption{Keywords from English Tweets}
    \footnote{Image tool credit \url{https://worditout.com/word-cloud/create}.}
    \label{fig:34}
\end{figure}

\begin{figure}[H]
    \centering
        \includegraphics[scale=.8]{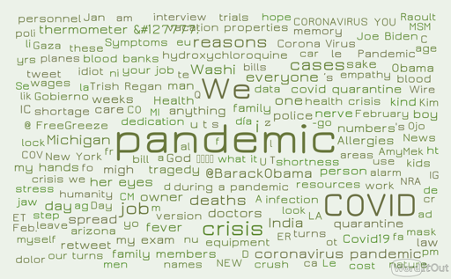}
    \caption{Word showing keywords from English tweets.}
    \label{fig:35}
\end{figure}

\begin{figure}[H]
    \centering
        \includegraphics[scale=.8]{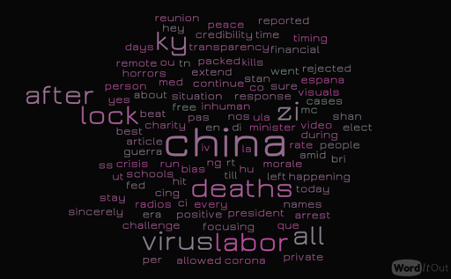}
    \caption{Word showing keywords from multiple-language tweets: a lexicon-based method.}
    \label{fig:36}
\end{figure}

\noindent C. \textbf{The reason why people feel sad or fear}: \\ Figure \ref{fig:34} shows some found keywords from English tweets. Figure \ref{fig:35} and \ref{fig:36} show multilingual keywords which contain English, Spanish, German, Portuguese, Japanese and Chinese based on different models. 

\begin{figure}[H]
    \centering
        \includegraphics[scale=.3]{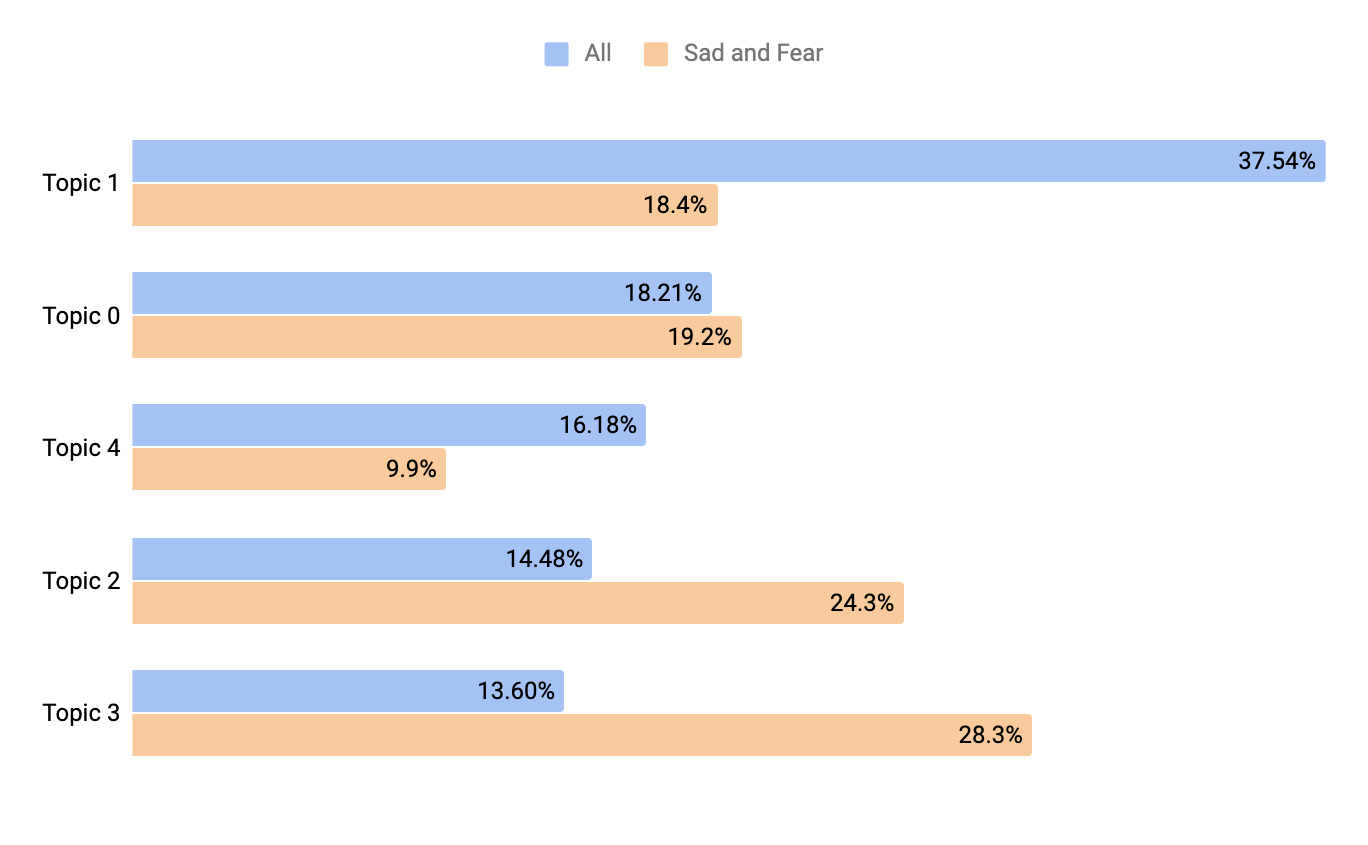}
    \caption{Topic Distribution}
    \label{fig:37}
\end{figure}

\noindent D. \textbf{The topics of Tweets}:\\
We set the number of topics to be 5, and did detailed analysis on the data from April 7th, 2020. The topics are listed as the following: 
\bigbreak
Topic 0: \texttt{Covid19 testing, deaths cases, positive cases}
\bigbreak
Topic 1: \texttt{President Trump, government, federal affairs}
\bigbreak
Topic 2: \texttt{lockdown, stay at home, physical distancing}
\bigbreak
Topic 3: \texttt{(Spanish) pandemic, health conditions}
\bigbreak
Topic 4: \texttt{the peak, serious treatment, Boris Johnson}
\bigbreak
Figure \ref{fig:37} shows the distribution of the topics.  We choose the data on April 7th, and first we do inference on all the data, and show the ratio for each topic learned above (All). Then we do inference on the tweets that are only labeled as sadness or fear (Sad and Fear). And the following is the ratio of each topic learned. In general, the public may be worried about Topic 3 and 2, mainly, the pandemic and lockdown, which are making people stressed. 

\subsection{Emotion Analysis on Instagram}
\subsubsection{Key Takeaway}
In the beginning of April 2020, the number of posts with \#depression on Instagram doubled, which might reflect the rise in mental health awareness among Instagram users.

\subsubsection{Data Description/ Methodology}
We collected 71,737 posts  on Instagram \cite{Facebook} with \#depression from March 31 to April 5, 2020. During this period, the number of the posts are steadily increasing as below. Among the posts with location information, \#depression was mostly posted in the U.S., the U.K. and India. In those countries, users posted the hashtag during the local daytime. During the long quarantine, people might struggle to keep their mental conditions healthy and the trend of \#depression on Instagram might reflect their attitudes.

\subsubsection{Overall Analysis}
The number of hourly posts with \#depression doubled in a week from March 31 to April 5. Among the posts with location information, most posts were uploaded in the U.S. In all top three countries, the U.S., the U.K. and India, the hashtag was mainly posted in between afternoon and the evening.

\begin{figure}[H]
    \centering
        \includegraphics[scale=.8]{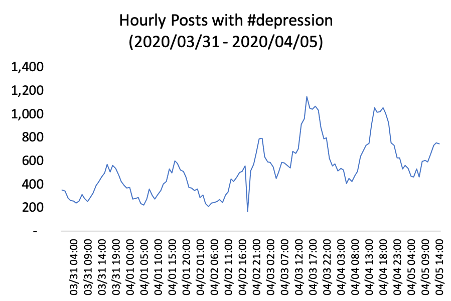}
    \caption{Hourly Posts with \#depression}
    \label{fig:38}
\end{figure}

\begin{figure}[H]
    \centering
        \includegraphics[scale=.85]{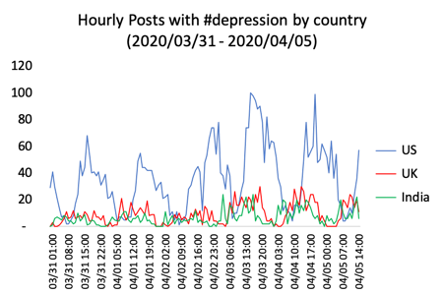}
    \caption{Hourly Posts with \#depression by Country}
    \label{fig:39}
\end{figure}

\bigbreak
\bigbreak

\section{Behaviour Changes}
\subsection{Analysis of Social Behavioral Changes on Instagram }
\subsubsection{Key Takeaway}
The number of posts with \#cough started increasing in mid of March, several days before the stay-at-home orders in some countries. Some governments could take their initial response earlier than they actually did based on the increase in the number of the behavioral changes on the social networking service platform. Other hashtags related to users behavioral changes, including \#mask, \#facemask, \#stayalive, started increasing in mid of March.

\subsubsection{Data Description/ Methodology}
We analyzed basic hashtags such as \#covid19 and \#coronavirus; hashtags related to medical supplies \#mask and \#facemask; \#ClapBecauseWeCare, a daily event people cheer medical professionals working on the frontline; and a hashtag that support others such as\#stayalive.

\subsubsection{Overall Analysis}
\begin{figure}[H]
    \centering
        \includegraphics[scale=.8]{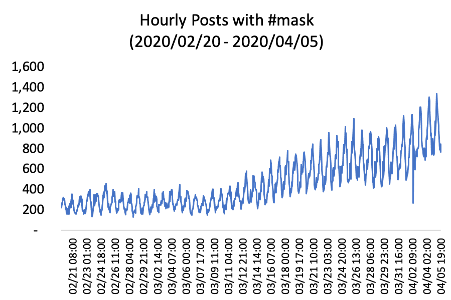}
    \caption{Hourly Posts with \#mask}
    \label{fig:40}
\end{figure}

\begin{figure}[H]
    \centering
        \includegraphics[scale=.8]{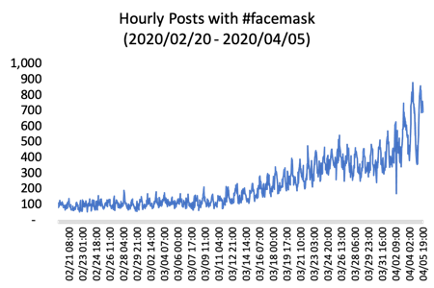}
    \caption{Hourly Posts with \#facemask}
    \label{fig:41}
\end{figure}

We have collected 513,712 posts with \#mask and 251,452 posts with \#facemask since February 20, 2020. Figure \ref{fig:40} and \ref{fig:41}show that the two hashtags have been steadily increasing since around March 11, 2020, a few weeks earlier than some lockdown announcements in Europe and stay-at-home orders in the U.S. People might have considered how to protect themselves amid the pandemic before their governments imposed strict prohibitions on residents.

In March 2020, \#cough was posted 19,983 times, 4 times more than that in February. The peak was March 15 in the U.S., March 20 in the U.K., March 21 in India. These dates are a few days earlier than the dates of stay-at-home orders in those countries, such as shelter-in-place order at San Francisco on March 16, the announcement of lockdown by the U.K. prime minister on March 23 and the statement of a complete nationwide lockdown by the prime minister in India on March 24. This implies that citizens had started to recognize the symptom a few days before the governments announced official statements. The governments could take their initial response to the pandemic earlier based on the increase in the number of \#cough around March 9.

\begin{figure}[H]
    \centering
        \includegraphics[scale=.8]{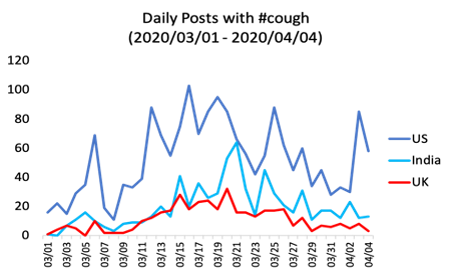}
    \caption{Daily Posts with \#cough}
    \label{fig:42}
\end{figure}

“Clap Because We Care” initiative encourages people around the world to make noise by singing or clapping from their home in order to show support for medical staff working on the frontlines of the COVID 19 situation. The initiative originally started in Wuhan, China in January, and has called citizen’s participation In Boston and New York City, at 7 p.m. on Fridays. With 4,033 posts after March 27, 2020, the graph below clears that some palse have been observed in the U.S. and the U.K. On Friday, March 27, approximately 180 \#ClapBecauseWeCare posts were uploaded followed by more than 100 posts at U.K. on Friday, April 2. These movements might spread to other countries and cities in the next few weeks.

\begin{figure}[H]
    \centering
        \includegraphics[scale=.8]{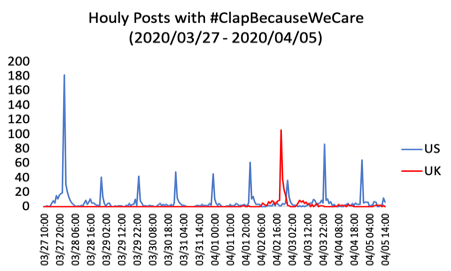}
    \caption{Hourly Posts with \#ClapBecauseWeCare}
    \label{fig:43}
\end{figure}

We analyzed 38,979 posts with \#stayalive from January 1 to April 6 in 2020. The posts have been gradually increasing since middle March 2020, when COVID-19 deaths in the world exceeded 10,000 and some states in America including California and New York declared emergency. The number of posts with the hashtag in March 2020 was twice more than that in February. This suggests that Instagram users started to take care of themselves and their friends by recognizing the severity of the situation through 10,000 death milestones.

\begin{figure}[H]
    \centering
        \includegraphics[scale=.8]{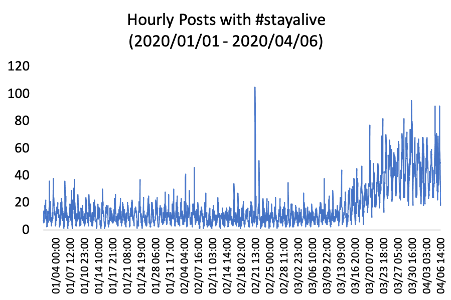}
    \caption{Hourly Posts with \#stayalive}
    \label{fig:44}
\end{figure}

\subsection{Perception and Resulting Behavioral Changes in Japan}
\subsubsection{Key Takeaway}
People’s perception regarding COVID-19 varies depending on the number of infected cases in the local community. Also, people’’s behavior has some indicative signal for the future spreading the disease.

\subsubsection{Data Description/ Methodology}
We analyzed the data of the national survey concerning COVID-19 news and resulting behavior changes on March 7 to 9 provided by Survey Research Center Co. Ltd. 100 responders were randomly selected from each prefecture from an approximate national pool of 2 million panelists. We also used national data regarding confirmed COVID-19 cases from March 6th (the day before the survey was conducted).

\subsubsection{Overall Analysis}
\begin{figure}[H]
    \centering
        \includegraphics[scale=.25]{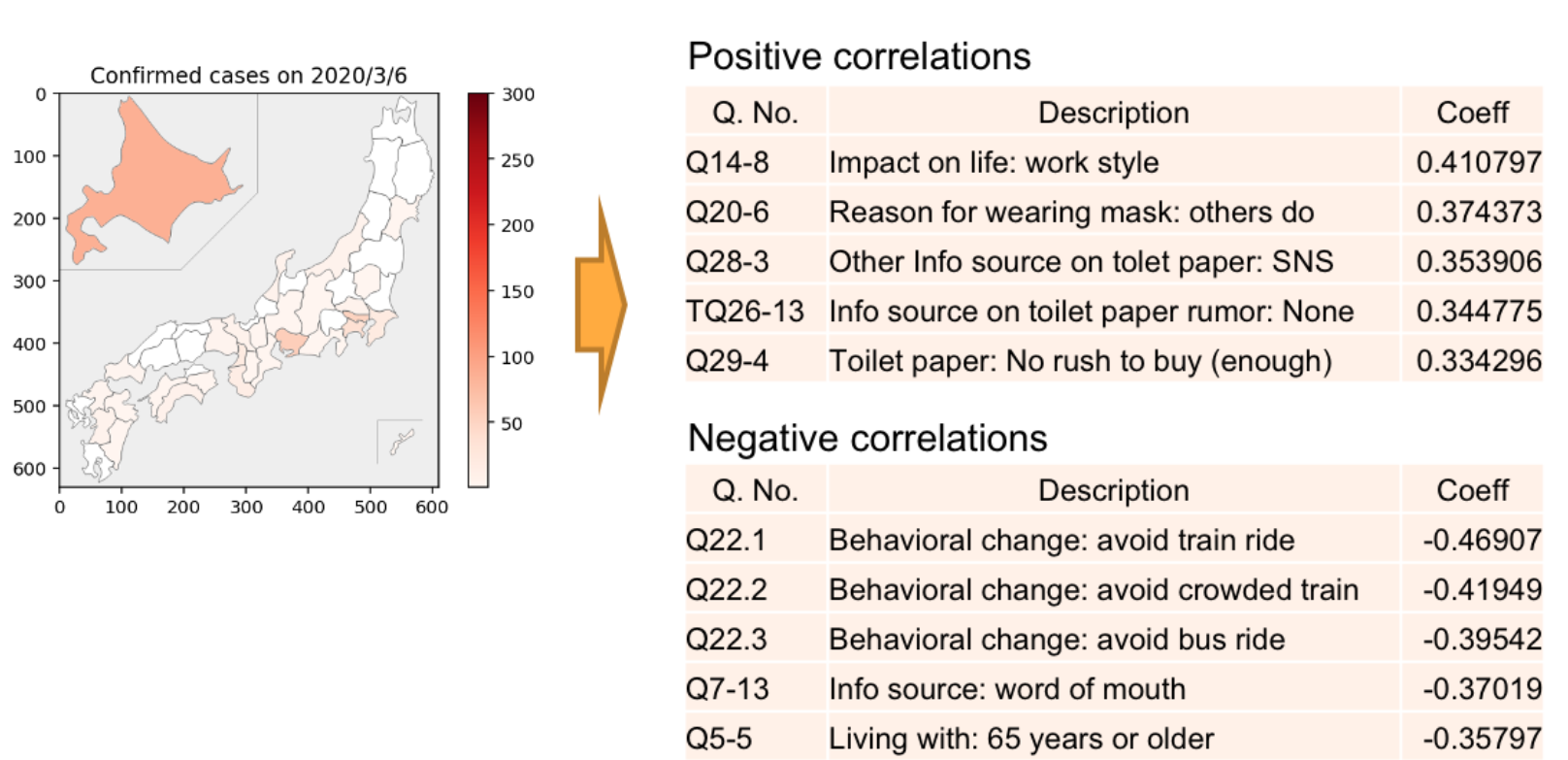}
    \caption{Correlation between confirmed cases before the survey (left) and survey responses (right)}
    \label{fig:45}
\end{figure}

\begin{figure}[H]
    \centering
        \includegraphics[scale=.25]{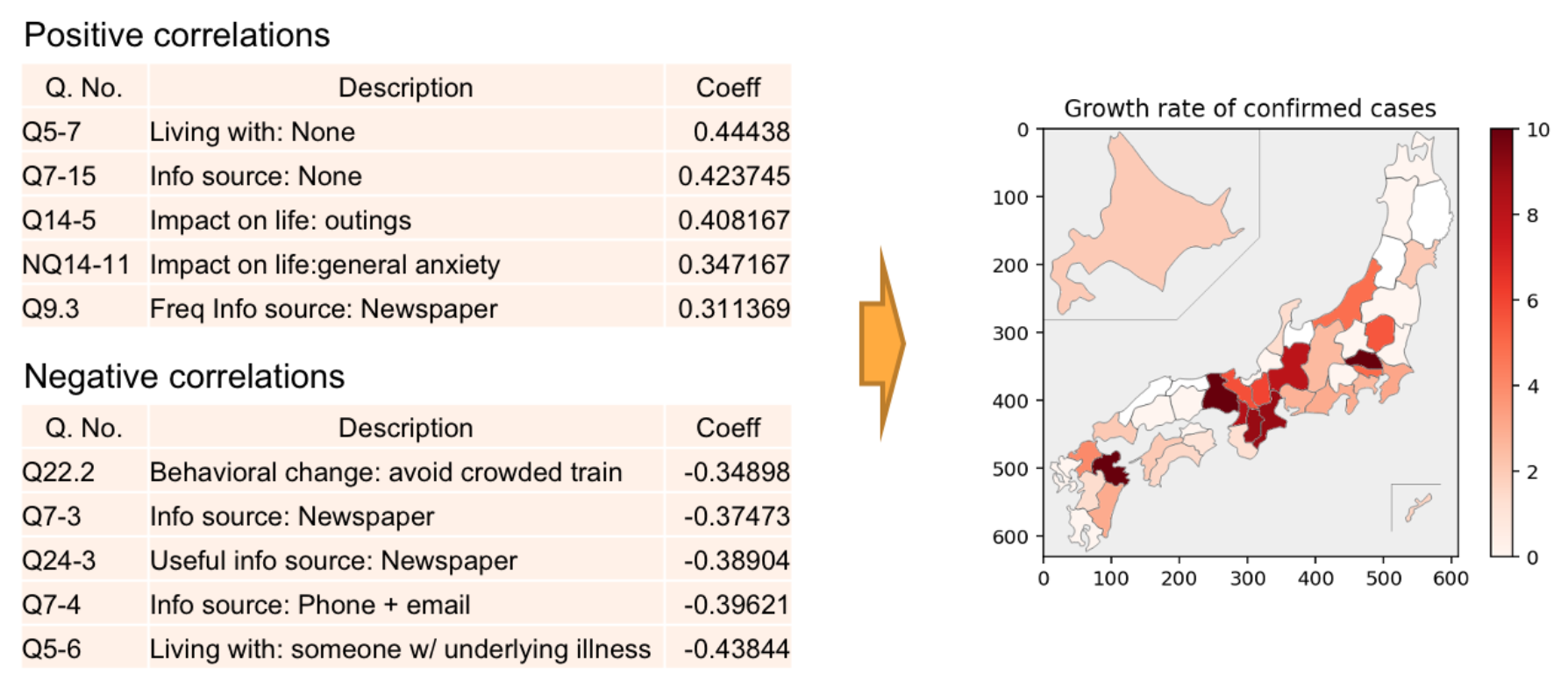}
    \caption{Correlations between the survey responses (left) and the growth rate of the reported confirmed cases on March 27th}
    \label{fig:46}
\end{figure}

In densely affected areas higher levels of concern regarding the impact of COVID-19 on everyday work (Q14-8) were observed. In areas with lower numbers of confirmed cases, commuting behavior was shifted to avoid public transportation.
\bigbreak
We also considered how people’s perception and behavior may impact the spread of the disease. Since there is some delay between an infection and its appearance in the official statistics, we took the growth rate of the reported confirmed cases on March 27th as the indicator of the rate of the spread. Figure \ref{fig:46} shows the strong positive and negative correlations.

Prefectures with higher numbers of single households and people that have less interest in current affairs show higher growth rates of confirmed cases. 

\bigbreak
\bigbreak

\section{Economics Analysis}
\subsection{Unemployment Impact in Catalonia}

\subsubsection{Key Takeaway}

Analysis of unemployment and temporal unemployment regulation files (Expediente de Regulación Temporal de Empleo or "ERTE" in Spanish) of Catalonia region show how unemployment has doubled in less than three months. Part of this unemployment is still to be defined, as it is currently under the figure of temporal unemployment. So far, solid unemployment has only grown 2.3\%. The upcoming months will be crucial in this regard, showing at what rate do ERTEs succeed in recovering workplaces. 
\bigbreak
The mobility analysis of Barcelona seemed to indicate that society assimilated the importance and severity of the situation during the second week of lockdown. This may be supported by this work, which indicates that this same week hosted most of the layoffs. This is however, before the toughest part of the lockdown (during the first two weeks, travel to work was allowed). In this regard, the hard lockdown seemed to have little further effect on unemployment.

\subsubsection{Data Description/ Methodology}
For this analysis we used the public ERTOs data from the \textit{Generalitat of Catalunya}\footnote{\url{https://analisi.transparenciacatalunya.cat/Treball/Evoluci-di-ria-dels-Expedients-de-Regulaci-Tempora/atmi-6snp}} and the public data of unemployment of \textit{SEPE}\footnote{\url{https://www.sepe.es/HomeSepe/que-es-el-sepe/estadisticas/datos-avance/datos.html}}.

\subsubsection{Overall Analysis}
\begin{figure}[H]
    \centering
        \includegraphics[scale=0.47]{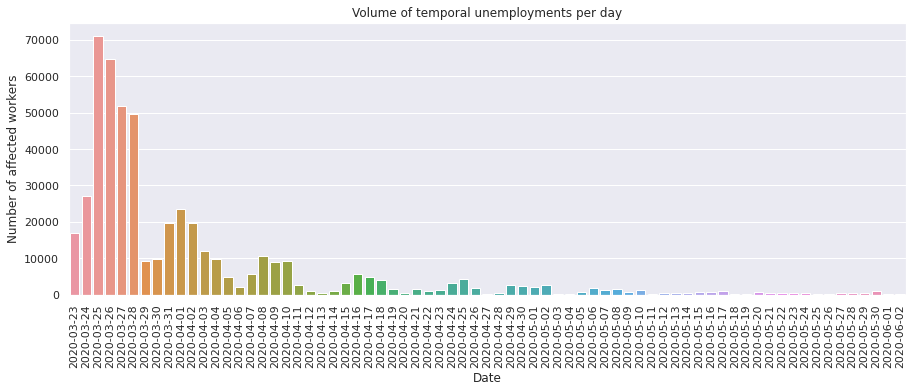}
    \caption{Volume of workers affected by temporal unemployment on each day}
    \label{fig:ertos_cat}
\end{figure}

During the lockdown, the Spanish government promoted a temporal unemployment fiscal figure (ERTE) under which companies can temporarily layoff workers. In this period, 70\% of the worker's salary is paid by the government, and the company may complement the rest. The purpose of this measure is to maximize the number of workplaces which are restored after the economic lockdown. The success of this measure will have a great impact on the duration of the economic side-effects of the pandemic. We analysed the effect of this measure in Catalonia, one of the most populated Autonomous Communities of Spain. Catalonia includes the city of Barcelona and close to 3.4M workers. Industrial activity represents nearly 21\% of the Catalan GDP, while tourism accounts for 12\%.
\bigbreak
First in Figure \ref{fig:ertos_cat} we plot the number of temporal unemployments per day. That is number of affected workers on each day. For context, the Spanish Government announced a state of emergency, and implemented a lockdown for the whole population on March 14th.  This lockdown was reinforced on March 29th to total mobility restrictions, with the only exception of essential services. 

\begin{figure}[H]
    \centering
        \includegraphics[scale=0.75]{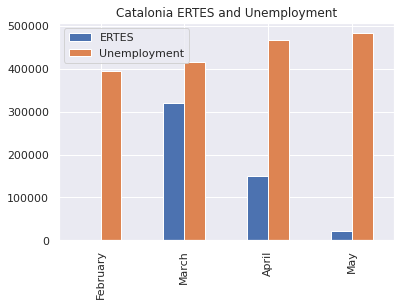}
    \caption{Comparison between general and temporal unemployment in the last months in Catalonia}
    \label{fig:ertos_unenployment_cat}
\end{figure}

As seen in Figure \ref{fig:ertos_cat}, most of the layoffs happened during the second week of the pandemic, before the the total lockdown. This indicates that, for the case of Catalonia, the mild lockdown and the hard lockdown have similar economic effects in terms of unemployment. On the worst day, 25th of March, 2\% of Catalan workers were fired. By the end of May, the total number of workers affected by temporal unemployment in Catalonia was 491,789.
\bigbreak
Next, in Figure \ref{fig:ertos_unenployment_cat}, we compare the unemployment volume with the volume of ERTEs per month. Considering the difference between the ERTEs volume (over 300K on the worse day) and the growth of unemployment (less than 100K accumulated), ERTEs seem to many of the layoffs, mitigating the growth of unemployment at least temporally. The behavior of unemployment in the coming months, as ERTEs expire, will provide the real measure on the effectiveness of ERTEs.

Finally, Table \ref{tab:ertos_table} shows the total number of unemployments and ERTEs. Unemployment in February was 10.4\%, and rose to 12.7\% in May. Meanwhile, accumulated ERTEs affected 12.9\% of workers. Combined, regular unemployment and ERTEs account for a 25.6\% of all workplaces.

\begin{table}[]
    \centering
    \begin{tabular}{lcccc}
    \toprule
    {} &   ERTEs & Unemployment &    ERTEs  \% & Unemployment  \% \\
    \midrule
    February &     --- &       395,214 &         --- &      10.4004 \\
    March    &  320,582 &       417,047 &   8.43637 &        10.9749 \\
    April    &  149,457 &       467,810 &   3.93308 &        12.3108 \\
    May      &   21,750 &       483,149 &   0.57236 &        12.7144 \\
    \bottomrule
    \end{tabular}

    \caption{Total number and percentage (with regard to the working population of 3.8M) of the two kinds of unemployment.}
    \label{tab:ertos_table}
    
\end{table}
\medskip

\clearpage

\bibliography{references}

\end{document}